\documentstyle[11pt]{article}

\newtheorem{THEOREM}{Theorem}[section]
\newenvironment{theorem}{\begin{THEOREM} \hspace{-.85em} {\bf :} }%
                        {\end{THEOREM}}
\newtheorem{LEMMA}[THEOREM]{Lemma}
\newenvironment{lemma}{\begin{LEMMA} \hspace{-.85em} {\bf :} }%
                      {\end{LEMMA}}
\newtheorem{COROLLARY}[THEOREM]{Corollary}
\newenvironment{corollary}{\begin{COROLLARY} \hspace{-.85em} {\bf :} }%
                          {\end{COROLLARY}}
\newtheorem{PROPOSITION}[THEOREM]{Proposition}
\newenvironment{proposition}{\begin{PROPOSITION} \hspace{-.85em} {\bf :} }%
                            {\end{PROPOSITION}}
\newtheorem{DEFINITION}[THEOREM]{Definition}
\newenvironment{definition}{\begin{DEFINITION} \hspace{-.85em} {\bf :} \rm}%
                            {\end{DEFINITION}}
\newtheorem{CLAIM}[THEOREM]{Claim}
\newenvironment{claim}{\begin{CLAIM} \hspace{-.85em} {\bf :} \rm}%
                            {\end{CLAIM}}
\newtheorem{EXAMPLE}[THEOREM]{Example}
\newenvironment{example}{\begin{EXAMPLE} \hspace{-.85em} {\bf :} \rm}%
                            {\end{EXAMPLE}}
\newtheorem{REMARK}[THEOREM]{Remark}
\newenvironment{remark}{\begin{REMARK} \hspace{-.85em} {\bf :} \rm}%
                            {\end{REMARK}}
 
\newcommand{\thm}{\begin{theorem}}
\newcommand{\lem}{\begin{lemma}}
\newcommand{\pro}{\begin{proposition}}
\newcommand{\dfn}{\begin{definition}}
\newcommand{\rem}{\begin{remark}}
\newcommand{\xam}{\begin{example}}
\newcommand{\cor}{\begin{corollary}}
\newcommand{\prf}{\noindent{\bf Proof:} }
\newcommand{\ethm}{\end{theorem}}
\newcommand{\elem}{\end{lemma}}
\newcommand{\epro}{\end{proposition}}
\newcommand{\edfn}{\bbox\end{definition}}
\newcommand{\erem}{\bbox\end{remark}}
\newcommand{\exam}{\bbox\end{example}}
\newcommand{\ecor}{\end{corollary}}
\newcommand{\eprf}{\bbox\vspace{0.1in}}
\newcommand{\beqn}{\begin{equation}}
\newcommand{\eeqn}{\end{equation}}

\newcommand{\bbox}{\vrule height7pt width4pt depth1pt}

\newcommand{\clm}{\begin{claim}}
\newcommand{\eclm}{\end{claim}}

\newcommand{\sat}{\models}

\newcommand{\rimp}{\Rightarrow}

\newcommand{\dimp}{\Leftrightarrow}

\newcommand{\band}{\bigwedge}
\newcommand{\union}{\cup}
\newcommand{\inter}{\cap}

\renewcommand{\phi}{\varphi}

\newcommand{\A}{{\cal A}}

\newcommand{\D}{{\cal D}}
\newcommand{\E}{{\cal E}}
\newcommand{\F}{{\cal F}}
\newcommand{\G}{{\cal G}}
\newcommand{\I}{{\cal I}}
\newcommand{\J}{{\cal J}}
\newcommand{\K}{{\cal K}}
\newcommand{\M}{{\cal M}}

\newcommand{\R}{{\cal R}}
\newcommand{\T}{{\cal T}}

\newcommand{\ie}{i.e.,~}

\newcommand{\respc}{resp.,\ }

\newcommand{\ol}{\setlength{\itemsep}{0pt}\begin{enumerate}}
\newcommand{\eol}{\end{enumerate}\setlength{\itemsep}{-\parsep}}
\newcommand{\ul}{\setlength{\itemsep}{0pt}\begin{itemize}}
\newcommand{\dl}{\setlength{\itemsep}{0pt}\begin{description}}
\newcommand{\edl}{\end{description}\setlength{\itemsep}{-\parsep}}
\newcommand{\eul}{\end{itemize}\setlength{\itemsep}{-\parsep}}

\newcommand\eqdef{=_{\rm def}}

\newcommand{\false}{\mbox{{\it false}}}
 
\newcommand{\Kaxc}{{\rm K}_n^C}

\newcommand{\Taxc}{{\rm T}_n^C}

\newcommand{\fouraxc}{{\rm S4}_n^C}

\newcommand{\fiveaxc}{{\rm S5}_n^C}

\newcommand{\Daxc}{{\rm KD45}_n^C}

\newcommand{\MP}{{\cal M}_n}
\newcommand{\MPr}{{\cal M}_n^r}
\newcommand{\MPrt}{\M_n^{\mbox{\scriptsize{{\it rt}}}}}
\newcommand{\MPrst}{\M_n^{\mbox{\scriptsize{{\it rst}}}}}
\newcommand{\MPelt}{\M_n^{\mbox{\scriptsize{{\it elt}}}}}

\newcommand{\commentout}[1]{}

\newcommand{\bi}{\begin{itemize}}
\newcommand{\ei}{\end{itemize}}
\newcommand{\be}{\begin{enumerate}}
\newcommand{\ee}{\end{enumerate}}

\renewcommand{\topfraction}{.9}
\renewcommand{\textfraction}{.1}
\renewcommand{\Kaxc}{{\rm K}_\G^C}
\newcommand{\KaxcG}{{\rm K}_{\G_\phi}^C}
\newcommand{\Kaxcp}{{\rm K}_{\G'}^C}
\renewcommand{\Taxc}{{\rm T}_\G^C}
\newcommand{\Taxcp}{({\rm T}_\G^C)^+}
\renewcommand{\fouraxc}{{\rm S4}_\G^C}
\newcommand{\fouraxcp}{({\rm S4}_\G^C)^{\A_1+\A_2}}
\renewcommand{\fiveaxc}{{\rm S5}_\G^C}
\newcommand{\fiveaxcp}{({\rm S5}_\G^C)^{\A_1+\A_2}}
\renewcommand{\Daxc}{{\rm KD45}_\G^C}
\newcommand{\Daxcp}{({\rm KD45}_\G^C)^{\A_1+\A_2}}
\renewcommand{\MP}{{\M_\A}}
\renewcommand{\MPr}{\M_\A^{\mbox{\scriptsize{{\it r}}}}}
\renewcommand{\MPrt}{\M_\A^{\mbox{\scriptsize{{\it rt}}}}}
\renewcommand{\MPrst}{\M_\A^{\mbox{\scriptsize{{\it rst}}}}}
\renewcommand{\MPelt}{\M_\A^{\mbox{\scriptsize{{\it elt}}}}}
\newcommand{\MPrth}{\M_{\A_1+\A_2}^{\mbox{\scriptsize{{\it rt}}}}}
\newcommand{\MPrsth}{\M_{\A_1+\A_2}^{\mbox{\scriptsize{{\it rst}}}}}
\newcommand{\MPelth}{\M_{\A_1+\A_2}^{\mbox{\scriptsize{{\it elt}}}}}
\newcommand{\Frst}{\F_\A^{\mbox{\scriptsize{{\it rst}}}}}
\newcommand{\LGC}{{\cal L}_{\G}^C}
\newcommand{\LGE}{{\cal L}_{\G}^E}
\newcommand{\LGp}{{\cal L}_{\G'}^C}
\newcommand{\Sub}{\mbox{\it Sub}}
\newcommand{\ESub}{\mbox{\it ESub}}
\renewcommand{\H}{{\cal H}}
\renewcommand{\L}{{\cal L}}
\newcommand{\hJ}{\widehat{{\cal J}}}
\newcommand{\tJ}{\widetilde{{\cal J}}}
\newcommand{\hG}{\widehat{{\cal G}}}
\newcommand{\tG}{\widetilde{{\cal G}}}
\newcommand{\Rmax}{\R(\G_\phi)}
\newcommand{\sigmak}{\sigma_1}
\newcommand{\sigmart}{\sigma_2}
\newcommand{\sigmarst}{\sigma_3}
\newcommand{\sigmaelt}{\sigma_4}
\newcommand{\sigmarstp}{\sigma_3'}
\newcommand{\tauk}{\tau_1}
\newcommand{\taurt}{\tau_2}
\newcommand{\taurst}{\tau_3}
\newcommand{\tauelt}{\tau_4}
\newcommand{\citeyear}{\cite}
\newcommand{\oKi}{\overline{K_i}}

\begin{document}

\begin{titlepage}
\title{Reasoning About Common Knowledge with Infinitely Many Agents}
\author{Joseph Y. Halpern%
\thanks{Supported in part by NSF under grant IRI-96-25901.}\\
Computer Science Department\\
Cornell University\\
halpern@cs.cornell.edu\\
\addtocounter{footnote}{2}
\and
Richard A. Shore%
\thanks{Supported in part by NSF Grant DMS-9802843
and DOD MURI Grant SA1515PG.}\\
Mathematics Department\\
Cornell University\\
shore@math.cornell.edu}
\date{\today}

\maketitle
\thispagestyle{empty}

\begin{abstract}
Complete axiomatizations and exponential-time decision procedures are
provided for reasoning about knowledge and common knowledge when there
are infinitely many agents.  The results show that reasoning about knowledge
and common knowledge with infinitely many agents is no harder than when
there are finitely many agents, provided that we can check the
cardinality of certain set differences $G - G'$, where $G$ and $G'$ are
sets of agents.  Since our complexity results are independent of the
cardinality of the sets $G$ involved,
they represent improvements over the
previous results even with the sets of agents involved are finite.
Moreover, our results make clear the extent to which issues of
complexity and completeness depend on how the sets of
agents involved are represented.
\end{abstract}

\end{titlepage}


\section{Introduction}
Reasoning about knowledge and common knowledge has been shown to be
widely applicable in distributed computing, AI, and game theory.  (See
\cite{FHMV} for numerous examples.)
Complete axioms for reasoning about knowledge and common knowledge are
well known in the case of a fixed finite set of agents.  However, in
many applications, the set of agents is not known in advance and
has no {\em a priori\/} upper bound (think of software agents on the web
or nodes on the Internet, for example); it is often easiest to model the
set of agents
as an infinite set.
Infinite sets of agents also arise in game theory and economics
(where
reasoning about knowledge and common knowledge is quite standard; see,
for example, \cite{Au,Gea94}).
{F}or example, when analyzing a game played with two teams,
we may well want to say that everyone on team 1 knows that
everyone on team 2 knows some fact $p$, or that it is common knowledge
among the agents
on
team 1 that $p$ is common knowledge among the agents
on
team 2.  We would want to say this even if the teams consist of
infinitely many agents.
Since economies are often modeled
as consisting of infinitely many (even uncountably many) agents, this
type of situation arises when economies are viewed as teams in a
game.

The logics for reasoning about the knowledge of groups of agents contain
modal operators $K_i$ (where $K_i \phi$ is read ``agent $i$ knows
$\phi$''), $E_G$ (where $E_G \phi$ is read ``everyone in group $G$ knows
$\phi$''), and $C_G$ (where $C_G \phi$ is read ``$\phi$ is common
knowledge among group $G$'').
The operators $E_G$ and $C_G$ make perfect sense even if we allow the
sets $G$ to be infinite---their semantic definitions remain unchanged.
If the set of agents is finite, so that,
in particular, $G$ is finite, there is a simple axiom connecting $E_G
\phi$ to $K_i \phi$, namely,
$E_G \phi \dimp \land_{i \in G} K_i \phi$.
Once we allow infinite groups $G$ of agents, there is no
obvious analogue for this axiom.  Nevertheless, in this paper,
we show that there exist natural sound and complete axiomatizations for
reasoning about knowledge and
common knowledge even if there are infinitely many agents.

It is also well known that if there are finitely many agents, then there
is a decision procedure that decides if a formula $\phi$ is satisfiable
(or valid)
that runs in time exponential in $|\phi|$, where $\phi$ is the length of
the formula viewed as a string of symbols.
We prove a similar result for a language with
infinitely
many
agents. However,
two issues arise (that, in fact, are also relevant
even if there are only finitely many agents, although they have not been
considered before):
\begin{itemize}
\item


In the statement
of the complexity result in \cite{FHMV}, $E_G$ and $C_G$ are both viewed
as having length $2 + 2|G|$
(where $|G|$ is the cardinality of $G$).
Clearly we cannot use this definition here if we want to get interesting
complexity results,
since $|G|$ may be infinite.  Even if we restrict
our attention
to finite sets $G$,
we would like a decision procedure that treats these sets in a uniform
way,
independently
of their cardinality.
Here we view $E_G$ as having
length 1 and $C_G$ as having length 3, independent of the cardinality of
$G$.
(See, for example, the proof of Proposition~\ref{Shore}
for the role of
independence and the definition of $Sub(\phi)$ in the proof
of Theorem~\ref{dec}
for an indication as to why $C_G$ has length
3 rather than $1$.)
Even with this definition of length, we prove that the complexity
of the satisfiability problem is still essentially exponential time.
(We discuss below what ``essentially'' means.)
Thus our results improve previously-known results even if there are
only finitely many agents.
\item

In the earlier proofs, it is implicitly
assumed
that the sets
$G$ are presented in such a way that there is no difficulty in testing
membership in $G$.   As we show here, in order to decide if certain
formulas are
satisfiable, we need to be able to test if certain subsets of agents of
the form $G_0 - (G_1 \union \ldots \union G_k)$ are empty, where $G_0,
\ldots, G_k$ are sets
of agents.
In fact, if we are interested in a notion of knowledge that
satisfies {\em positive introspection\/}---that is, if agent $i$ knows
$\phi$, then she knows that she knows it---then we also
must be able to check whether such subsets are singletons.
And if we are interested in a notion of knowledge that
satisfies {\em negative introspection\/}---that is, if agent $i$
does not know
$\phi$, then she knows that she does not know it---then
we
must be able to
check whether such subsets have cardinality
$m$, for certain finite $m$. The difficulty of deciding these
questions depends in part on how $G_0,
\ldots, G_k$ are presented and which sets of agents we can talk about in
the language.  For example, if $G_0, \ldots, G_k$ are
recursive sets, deciding if $G_0 - (G_1 \union \ldots \union G_k)$ is
nonempty may not even be recursive.  Here, we
provide a
decision procedure for satisfiability that runs in time exponential in
$|\phi|$ provided
that we have oracles for testing appropriate properties of sets of the
form $G_0 - (G_1 \union \ldots \union G_k)$.  Moreover, we show that
any decision procedure
must
be able to answer the questions
we ask.
In fact, we actually prove a
stronger result, providing a tight bound
on the complexity of deciding satisfiability that takes into account the
complexity of answering questions about the cardinality of $G_0 - (G_1
\union \ldots \union G_k)$.

Again, this issue is of significance even if there are only finitely
many agents.  For example, in the SDSI approach to security \cite{RL96},
there are names, which can be viewed as representing sets of agents.
SDSI provides a (nondeterministic) algorithm for computing
the set of agents represented by a name.  If we want to make statements
such as ``every agent represented by name ${\tt n}$ knows $\phi$''
(statements that we believe will be useful in reasoning about
security \cite{HM99,HMS}) then the results of this paper show that to
decide validity in the resulting logic, we need more than just an
algorithm for
resolving the agents represented by a
given name.  We also need
algorithms for resolving
which agents are
represented by one name and not
another.  More generally, if we assume that we have a separate language
for representing sets of agents, our results characterize the properties
of sets that we need to be able to decide in order to reason about the
group knowledge of these agents.
\end{itemize}

In the next section, we briefly review the syntax and semantics of
the logic of common knowledge.
In Section~\ref{simplified} we state the main results and prove them
under some simplifying assumptions that allow us to bring out the main
ideas of the proof.
We drop these assumptions in Section~\ref{proofs}, where we provide
the proofs of the full results.

\section{Syntax and Semantics: A Brief Review}\label{review}
\paragraph{Syntax:}  We start with a (possibly infinite) set $\A$ of
agents.  Let $\G$ be a set of nonempty subsets of $\A$.  (Note that we
do not require $\G$ to be closed under union, intersection, or
complementation; it can be an arbitrary collection of subsets.)
We get the language $\LGC(\Phi)$ by starting with a set $\Phi$ of
primitive propositions, and closing under $\land$, $\neg$, and the modal
operators $K_i$,
for
$i \in\A$, and $E_G, C_G$, for $G \in \G$.  Thus, if
$p, q
\in\Phi$, $i \in \A$, and $G, G' \in\G$, then $K_i C_G (p \land E_{G'}
q) \in \LGC(\Phi)$.
Let $\LGE$ be the sublanguage of
$\LGC$ that does not include the $C_G$ operators.
Let $|\phi|$ be the length of the formula viewed as a string of
symbols, where the modal operators $K_i$ and $E_G$ are counted as having
length 1 and $C_G$ is
counted as having length 3 (even if $G$ is an infinite set of agents)
and all primitive propositions are counted as having length 1.

In \cite{FHMV,HM2}, $\A$ is taken to be the set $\{1,\ldots,n\}$;
in \cite{HM2}, $\G$ is taken to be the singleton
$\{\{1,\ldots,n\}\}$
(so that we can only talk about every agent in $\A$ knowing $\phi$
and common knowledge among
the agents in $\A$), while in
\cite{FHMV}, $\G$ is taken to consist of all nonempty subsets of $\A$.

\paragraph{Semantics:}  As usual, formulas in $\LGC$
are either true or false at a world in a Kripke structure.  Formally, a
Kripke structure $M$ over $\A$ and $\Phi$
is a tuple $(S, \pi, \{\K_i: i \in \A\})$, where $S$ is a set of states
or possible worlds, $\pi$ associates with each state in $S$ a truth
assignment to the primitive propositions in $\Phi$ (so that $\pi(s):
\Phi \rightarrow \{{\bf true},{\bf false}\}$), and $\K_i$ is a binary
relation on $S$ for each agent $i \in \A$.  We occasionally write
$\K_i(s)$ for $\{t: (s,t) \in \K_i\}$.

We define the truth relation $\sat$ as follows:
\begin{description}
\item[]
$(M,s)\sat p$ (for $p\in\Phi$) iff $\pi(s)(p)={\bf true}$
\item[]
$(M,s)\sat\varphi\land \psi$  iff both  $(M,s)\sat \varphi$ and
$(M,s)\sat \psi$
\item[]
$(M,s)\sat \neg \varphi$ iff $(M,s)\not\sat \varphi$
\item[]
$(M,s)\sat K_i\varphi$  iff  $(M,t)\sat \varphi$ for all $t \in
\K_i(s)$
\item[]
$(M,s) \sat E_G \phi$ iff $(M,s) \sat K_i \phi$ for all $i \in G$
\item[]
$(M,s) \sat C_G \phi$ iff $(M,s) \sat E_G^k \phi$ for $k = 1, 2, 3,
\ldots$, where $E_G^k$ is defined inductively by taking
$E_G^1\phi \eqdef E_G \phi$ and $E_G^{k+1} \phi \eqdef
E_G E_G^k \phi$.
\end{description}
We say that
$t$ is {\em $G$-reachable from $s$ in $M$\/} if
there exist
$s_0, \ldots, s_k$ with $s = s_0$, $t= s_k$, and $(s_i, s_{i+1}) \in
\union_{i \in G}\K_i$.  For later use, we extend this
definition so that if $S' \subseteq S$, we say that $t$ is
{\em $G$-reachable from $s$ in $S'$\/} if $s_0, \ldots, s_k \in S'$.
The following characterization of common knowledge is well known
\cite{FHMV}.

\lem\label{G-reach} $(M,s) \sat C_G \phi$ iff $(M,t) \sat \phi$ for all
$t$
that
are $G$-reachable from $s$ in $M$.
\elem
Let   $\M_\A(\Phi)$\glossary{\glosmznap}
be the class of all Kripke
structures over $\A$ and $\Phi$ (with no restrictions on the
$\K_i$ relations).
We are also interested in
various subclasses of $\M_\A(\Phi)$, obtained by restricting the $\K_i$
relations.
In particular, we consider
$\MPr(\Phi)$, $\MPrt(\Phi)$, $\MPrst(\Phi)$, and
$\MPelt(\Phi)$,
the class of all structures over $\A$ and $\Phi$ where the $\K_i$
relations are reflexive (\respc reflexive
and transitive; reflexive, symmetric, and transitive; Euclidean,%
\footnote{Recall that a relation $R$ is Euclidean if $(s,t), (s,u) \in
R$ implies that $(t,u) \in R$.}
serial, and transitive).
For the remainder of this paper, we
take $\Phi$ to be fixed, and do not mention it, writing, for example
$\LGC$ and $\MP$ rather than $\LGC(\Phi)$ and $\MP(\Phi)$.

As usual, we define a formula to be {\em valid in a class $\M$\/} of
structures
if $(M,s) \sat \phi$ for all $M \in\M$ and all states $s$ in $M$;
similarly, $\phi$ is {\em satisfiable in $\M$\/} if $(M,s) \sat \phi$
for some $M \in \M$ and some $s$ in $M$.

\paragraph{Axioms:}
The following are the standard axioms
and rules
that have been considered for
knowledge;
They hold for all $i \in \A$.

\begin{description}
\item[Prop.]
All substitution instances of tautologies of propositional calculus.
\item[K1.]
$(K_i\varphi\land K_i(\varphi\rimp  \psi)) \rimp
 K_i\psi$.
\item[K2.] $K_i \phi \rimp \phi$.
\item[K3.] $\neg K_i \false$.
\item[K4.] $K_i \phi \rimp K_i K_i \phi$.
\item[K5.] $\neg K_i \phi \rimp K_i \neg K_i \phi$.
\item[MP.]
{F}rom $\varphi$ and $\varphi\rimp\psi$
infer $\psi$.
\item[KGen.]
{F}rom $\varphi$ infer $K_i\varphi$.

\end{description}
Technically, Prop and K1--K5 are axiom schemes, rather than single
axioms.
K1, for example, holds for all formulas $\phi$ and $\psi$.  A formula
such as $K_1 q \lor \neg K_1 q$ is an instance of axiom Prop
(since it is a substitution instance of the propositional tautology
$p \lor \neg p$, obtained by substituting $K_1 q$ for $p$).

We will be interested in the following axioms and rule
for reasoning about everyone knows,
which hold for all $G \in \G$.
\begin{description}
\item[E1.] $E_G \phi \rimp K_i \phi$ if $i \in G$.
\item[E2.] $(\land_{i \in \A'} K_i \phi \land \land_{G' \in \G'}
E_{G'} \phi)
\rimp E_{G} \phi$ if $\A'$ is a finite subset of $\A$, $\G'$ is a
finite subset of $\G$,
and $G \subseteq
(\A' \union (\union \G'))$.
\item[E3.]
$(E_G\varphi\land E_G(\varphi\rimp  \psi)) \rimp
E_G\psi$.
\item[E4.] $E_G (E_G \phi \rimp \phi)$. 
\item[E5.] $E_G \phi \rimp \phi$. 
\item[E6.] $\neg \phi \rimp E_G \neg E_G \phi$.
\item[E7.] From $\neg(\phi_1 \land \ldots \land \phi_k)$ infer
$\neg(E_{G_1} \phi_1 \land \ldots \land E_{G_k} \phi_k)$ if $G_1 \inter
\ldots \inter G_k \ne \emptyset$.
\item[EGen.] From $\varphi$ infer $E_G\varphi$. 
\end{description}

E2 can be viewed as a generalization of the axiom
$E_G \phi \rimp E_{G'} \phi$ if $G' \subseteq G$ (of which E1 is a
special case if we identify $K_i \phi$ with $E_{\{i\}} \phi$, as we
often do in the paper).  Essentially it says that if $K_i \phi$ hold for
all agents $i \in G$ (and perhaps some other agents $i \notin G$) then
$E_G \phi$ holds.  Since, if $G$ is infinite, we cannot write the
infinite conjunction of $K_i \phi$ for all $i \in G$, we approximate
as well as we can within the constraints of the language.  As long as
$E_{G'} \phi$ and $K_i \phi$ holds for sets $G'$ and agents $i$ whose
union contains $G$, then certainly $E_G \phi$ holds.

If $\A$ is finite (so that all the sets in $\G$ are finite)
we can simplify E1 and E2 to
\begin{description}
\item[E.] $E_G \phi \dimp \land_{i \in G} K_i \phi$.
\end{description}
It is easy to see that E follows from E1 and E2 (in the presence of
Prop and MP) and every instance of E1 and E2 follows from E if $\A$ is
finite.  E is used instead of E1 and E2 in \cite{FHMV,HM2}.  Note that
E2 is recursive iff deciding if $G - (\A' \union (\union \G')
) = \emptyset$ is recursive.
(We
determine precisely which such questions we must be able to
answer in Proposition~\ref{reduction}.)

E3 and EGen are the obvious analogues of K1 and KGen for $E_G$.  We do
not need them in the case that $\A$ is finite; it is easy to see
that they follow from K1, KGen, and E.  In the case that $\A$ is
infinite, however, they are necessary.

Axiom E4 is sound in $\MPr$, $\MPrt$, $\MPrst$, and $\MPelt$.
It is easy to see that E4 follows from K2, E1, and EGen, so will not be
needed in systems that  contain these axioms.  Moreover,
it is not hard to show that E4 follows from E1, E2, and K5 if the set of
agents is finite.  However, it does not follow from these axioms if the
set of agents is infinite.

Axiom E5 follows from K2 and E1.  Moreover, we use
it only in systems
that already include K2 and E1. Nevertheless, for
technical reasons, it is useful to
list it
separately.
Similarly, it is not hard to see that E7 is a derivable rule in any system
that includes Prop, MP, K1, K3, E1, E4, and EGen
(we prove this in Section~\ref{sec:elt}).  While we use E7 only in such
systems, like E5, it is useful to list it separately.



Axiom E6 (with $E_G$ replaced by $K_i$) is the standard axiom
used to characterize symmetric $\K_i$ relations \cite{FHMV}.  It follows
easily from K2, K5, E1, and E2 if $\A$ is finite.  However, like E4,
it must be
specifically included
if $\A$ is infinite.

{F}inally, we have the following well-known axiom and inference rule for
common knowledge:

\begin{description}
\item[C1.] $C_G \phi\rimp E_G(\phi \land C_G\phi$).
\item[RC1.]
{F}rom $\phi \rimp E_G(\psi \land \phi)$ infer $\phi
\rimp C_G\psi$.
\end{description}

Historically,
in the case of one agent,
the system with axioms and rules Prop, K1, MP, and KGen has been
called~K; adding K2 to K gives us T; adding K4 to T gives
us S4; adding K5 to S4 gives us S5; replacing K2 by K3 in S5 gives us
KD45. We use the subscript $\G$ to emphasize the fact that we are
considering systems with sets of agents coming from $\G$ rather than
only one
agent and the superscript $C$ to emphasize that we add E1--E3, EGen, C1,
and RC1 to the system.  In this way, we get the systems
$\Kaxc$\glossary{\gloskc},
$\Taxc$\glossary{\glostnc}, and
$\fouraxc$\glossary{\glosszec}.
Thus,
$\Kaxc$ consists of Prop, K1, MP, KGen, E1, E2, E3, EGen,
C1, and RC1; we get
$\fouraxc$
by adding
K2 and K4
to $\Kaxc$.
We get $\Daxc$\glossary{\gloskdefc} by adding K3--K5 and E4
to $\Kaxc$ and we get
$\fiveaxc$\glossary{\glosszfc} by adding
K2, K4, K5
and E6 to $\Kaxc$.

One of the two main results of this paper shows that each
of these axiom systems is sound and complete with respect to an
appropriate class of structures.  For example, $\Kaxc$ is a sound and
complete axiomatization
with respect to $\MP$
and $\fiveaxc$ is a sound and complete axiomatization with respect to
$\MPrst$.  In the case that $\A$ is finite, this result is well known
(see \cite{FHMV,HM2}---as mentioned earlier, E is used in the
axiomatization instead of E1--E3 and EGen).  What is perhaps surprising
is that
E1--E3 and EGen suffice even if $\A$ is infinite.  For example, suppose
that $\G$ just
consists of the singleton $\A$.  In that case, E2 becomes vacuous.
Thus, while the axioms force $E_\A \phi$ to imply
that each agent in $\A$ knows $\phi$, we have no way of expressing the
converse.
Indeed,
it is easy to construct a structure for the axioms with the
standard interpretations of all the $K_i$ relations but a
nonstandard one of $E_\A$, where all the agents in $\A$ know
$\phi$ and yet $E_\A \phi$ does not hold.
Consider, for example, a structure with a single state $s$ for the
language with an infinite set $\A$ of agents.
Suppose that every  primitive proposition $p$ is true at $s$,
$\K_i$ is empty for all $i \in \A$,
and
$K_i$ is interpreted in the usual way for all $i \in \A$ (so that $K_i
\phi$ is true at $s$ for all formulas $\phi$).
{F}or $E_\A$, however, we say
that $E_\A \phi$ holds at $s$ if and only if it is provable in, say,
$\Kaxc$. Of course, there are obviously standard models in which $E_\A
p$ does not hold and so (by the soundness of the
axioms for standard interpretations) $E_\A p$ is not provable. Thus,
in this interpretation, $E_\A p$ does not hold at $s$ while
$K_i p$ does for every $i \in \A$. Finally, it is clear that all the
axioms of $\Kaxc$ are true in this structure.
Similar examples can be given to
show
that E4 and
E6 do not follow from the specified other axioms when the set of agents
is infinite.

\section{The Main Results and a Proof in a Simplified Setting}
\label{simplified}

In this section, we state
the two main results of this paper---complete
axiomatizations and decision procedures.
We then provide a proof of a simpler version of these results that
illustrates some of the main ideas.  We first state the
completeness
results.

\thm\label{complete}  For formulas in the language $\LGC$:
\begin{enumerate}
\item[(a)]     $\Kaxc$  is a sound and complete axiomatization
with respect to $\MP$,
\item[(b)]     $\Taxc$  is a sound and complete axiomatization
with respect to $\MPr$,
\item[(c)]     $\fouraxc$  is a sound and complete axiomatization with
respect to $\MPrt$,
\item[(d)]     $\fiveaxc$  is a sound and complete axiomatization with
respect to $\MPrst$,
\item[(e)]     $\Daxc$  is a sound and complete axiomatization with respect
to $\MPelt$.
\end{enumerate}
\ethm

Before stating the
results
regarding complexity,
we first show that questions about certain facts regarding
sets of the form $G_0 - (G_1 \union \ldots \union G_k)$ are reducible to
satisfiability.
We are not just interested in sets of the form $G_0 - (G_1 \union \ldots
\union G_k)$ for $G_1, \ldots, G_k \in \G$.  For example, when dealing
with $\M^{rt}$, it turns out that we are interested in sets $H$ of this
form if $|H|=1$.  But if $H_1$ is such a set, then we are also
interested in sets of the form $H_2 = G_0 - (G_1 \union \ldots G_k
\union H_1)$.  And if $|H_2| = 1$, then we can also include $H_2$ in the
union, and so on.  The following definition makes this precise.

\dfn\label{Gm} Given a set $\J$ of subsets of $\A$ and
an integer $m \ge 1$, define a sequence $\J^m_0, \J^m_1,
\ldots$ of sets of subsets of
$\A$ inductively as follows.  Let $\J^m_0 = \J$.  Suppose that we have
defined $\J^m_0, \ldots, \J^m_{k}$.  Then
$\J^m_{k+1} = \J \union
\{G - \union \H:
G \in \J,\,
\H \subseteq
\J^m_{k},\,
\H \mbox{ finite}, \, |G - \union
\H| \le m\}$.
Let $\J^m = \union_i \J_i^m$; let $\hJ^m = \{G - \union \H: G \in \J, \H
\subseteq \J^m, \H$ finite$\}$.  For uniformity, we take $\hJ^0 =
\{G - \union \H: G \in \J, \H \subseteq \J, \H$ finite$\}$.
\edfn

Let $\J^*$ be the algebra generated by $\J$ (that is, the Boolean
combinations
of sets in $\J$).
It is useful to talk about the length of a description of various
sets in $\J^*$ (particularly those in $\hJ^m$ for some $m$).  Formally,
we assume we have a language
whose primitive objects
consist
of the elements of $\J$ and the symbols
$\union$ and $-$ (for set difference).  The length of a
description is then the number
of symbols of $\J$ that appear in it.  Notice that, in general, an
element of $\J^*$ may have several different descriptions.  We
are not always careful to distinguish a set from its description. (We
hope that the reader will be able to tell which is intended from
context.)  We use $l(G)$ to denote the length of
the description of $G \in \J^*$.

Let $\G_\A = \G \union \{\{i\}: i \in \A\}$.  Throughout the paper
(and, in particular, in the proof of the next proposition),
for ease of exposition, we identify
$E_{\{i\}}$ with $K_i$, for $i \in \A$
(which allows us to write $E_G$ for each $G \in \G_\A$).

\pro\label{reduction}
\begin{itemize}
\item[(a)] The question of whether $|G| > 0$ for $G \in \hG_\A^0$
is
reducible (in time linear in
$l(G)$) to the
satisfiability problem for the language $\LGE$
with respect to all of $\MP$, $\MPr$, $\MPrt$,
$\MPrst$, and $\MPelt$.
\item[(b)]
The
questions of whether $|G| >0$ and $|G| > 1$ for $G \in
\hG_\A^1$
are
each
reducible (in time linear in
$l(G)$) to the satisfiability
problem for the language $\LGE$
with respect to all of $\MPrt$, $\MPrst$, and $\MPelt$.
\item[(c)]
{F}or
all $m \ge 1$, the question of whether
$|G| > m$ for $G \in \hG_\A^m$ is reducible
(in time
linear in $l(G) + m$) to
the satisfiability problem for $\LGE$ with respect to $\MPrst$ and
$\MPelt$.
\item[(d)]
The
question of whether $|G_1 \inter \ldots \inter G_k| >
0$, for $G_1, \ldots, G_k \in \G_\A$
is reducible (in time linear in $k$) to the satisfiability problem
for $\LGE$ with respect to $\MPelt$.
\end{itemize}
\epro

\prf For part (a), suppose that $G \in \hG_\A^0$.  Thus, $G
= G_0 - (G_1 \union \ldots \union G_k)$
for some $G_0, \ldots, G_k \in \G_\A$.  Consider the formula
$\phi_a \eqdef
\neg E_{G_0}p \land E_{G_1} p \land \ldots \land E_{G_k} p $, where $p$
is a
primitive proposition. Clearly
$\phi_a$ is satisfiable in $\MP$, $\MPr$, $\MPrt$, $\MPrst$, or $\MPelt$
iff $|G_0 - (G_1 \union \ldots \union G_k)| > 0$.

{F}or part (b), given $G$,
we construct two formulas $\phi_{G,p}$
and $\psi_{G}$ with the following properties.
\begin{itemize}
\item $\phi_{G,p}$ is satisfiable iff $|G| > 0$.
\item If $(M,s) \sat \phi_{G,p}$, then
$(M,s) \sat \neg K_j p$ for some
$j \in G$.
\item $\psi_{G}$ is satisfiable iff $|G| > 1$.
\item $|\phi_{G,p}|$ and $|\psi_G|$ are both linear in $l(G)$.
\end{itemize}
This, of course, suffices to prove the result.

We construct the formulas $\phi_{G,p}$ by induction on the least $h$
such that $G =  G' - \union \H$ and $\H \subseteq
(\G_\A)^1_h$.
(We are here thinking of $G$ as specified by its description.)
If $\H \subseteq
(\G_\A)^1_0 = \G_A$, suppose that $\H = \{G_1, \ldots, G_k\}$.   Then we
take
$\phi_{G,p}$ to be $\neg E_{G'} p \land
E_{G_1} p \land \ldots \land E_{G_k}p$.  This clearly has the desired
properties.

Now suppose that
$\H \subseteq (\G_\A)^1_h$ for $h \ge 1$.
Without loss of generality, we can assume
that
$\H = \{G_1,
\ldots, G_{k'}, G_{k'+1} , \ldots , G_k\}$, where
$G_1, \ldots, G_{k'} \in
\G_\A$ and, for $j = k'+1, \ldots, k$,
$G_j \in (\G_\A)^1_h - \G_\A$ is of the
form $G_j' - \union \H_j$ with $G'_j \in \G_\A$, $\H_j \subseteq
(\G_\A)^1_{h-1}$, and $|G_j| = 1$.
Define $\phi_{G,p}$ as
$$\neg E_{G'} \neg (\neg p \land
\!\!\!\ \band_{j=k'+1}^k \!\!\!\!
\phi_{G_j,p_j}) \land
E_{G_1} p \land \ldots \land E_{G_{k'}} p \land
\!\!\! \band_{j=k'+1}^k\!\!\!\!
E_{G_j'} p_j,$$
where we assume that the sets of primitive propositions that appear in
$\phi_{G_j,p_j}$, $j= k'+1, \ldots, k$, are mutually exclusive and do
not include $p$.%
\footnote{Here we are implicitly assuming that the set of primitive
propositions is infinite, so that this can be done.  With more effort,
we can prove a similar result even if the set is finite, using the
techniques of \cite{Hal12}.}

Now suppose that $\phi_{G,p}$ is
true
at some state $s$ in 
$M \in \MPrt$.  Then for some $i \in G'$, we must have $(M,s)
\sat \neg K_i \neg (\neg p \land \band_{j=k'+1}^k \phi_{G_j,p_j})$.
We cannot have $i \in G_1 \union \ldots \union G_{k'}$, since $(M,s)
\sat E_{G_j} p$ for $j = 1, \ldots, k'$.  Nor can we have $i \in G_j$
for $j = k'+1, \ldots, k$.  For suppose that $G_j = \{i_j\}$, $j
\in \{k'+1,\ldots,k\}$.  Then $(M,s) \sat
\neg K_i \neg \phi_{G_j,p_j} \land E_{G_j'} p_j$,
which implies that
$(M,s) \sat
\neg K_i K_{i_j} p_j \land K_{i_j} p_j$.  Thus, we cannot have $i =
i_j$.  It follows that $G \ne \emptyset$.

Conversely, if $G  \ne
\emptyset$, we show that $\phi_{G,p}$ is satisfiable in $\MPrst$
(and hence also in $\MPrt$ and $\MPelt$).  We actually prove a stronger
result.  We show that if $G_1, \ldots, G_k$ are nonempty and the formulas
$\phi_{G_1,p_1}, \ldots, \phi_{G_k,p_k}$
involve disjoint sets of primitive propositions, then
$\phi_{G_1,p_1} \land \ldots \land \phi_{G_k,p_k}$ is satisfiable in
a structure in $\MPrst$ of a certain form.  To make this precise, suppose
that $M = (S,\pi, \{\K_i, i \in \A\})$, $s \in S$, $S'$ is
a set of states disjoint from $S$, and $s' \in S'$.  We say that $M$ is
{\em embedded
in
the structure $M' = (S \union S', \pi', \{\K_i',  \in \A\})$ at
$(s,s')$\/} if
\begin{enumerate}
\item $\pi'|_S = \pi$ and $\K_i'|_{S \times S} = \K_i$ for $i \in \A$,
\item if $(t,t') \in \K_i'$ for $t \in S$ and
$t' \in S'$, then $t=s$ and $t' = s'$.
\end{enumerate}
We show by induction on $h$ that if $G_1, \ldots, G_k \in
(\G_\A)^1_h$, $|G_i| > 0$ for $i = 1, \ldots, k$, and the formulas
$\phi_{G_1,p_1}, \ldots, \phi_{G_k,p_k}$
involve disjoint sets of primitive propositions, then for all $i_1,
\ldots, i_k$ such that $i_j \in G_j$, there exists a
structure $M \in \MPrst$ and a state $s$ in $M$ such that:
\begin{enumerate}
\item $(M,s) \sat \phi_{G_1,p_1} \land \ldots \land \phi_{G_k,p_k}$,
\item $\exists t_1, \dots, t_k$ such that $(s,t_j) \in \K_{i_j}$ and
$(M,t_j)\sat \neg p_j$,
\item $\K_i(s) = \{s\}$ for $i \notin \{i_1, \ldots, i_k\}$,
\item for all structures $M'$ and states $s'$ in $M'$ such that $M$ is
embedded in $M'$ at $(s,s')$ and
$(M',s') \sat
p_1 \land \ldots \land p_k$, we have that $(M',s) \sat \phi_{G_1,p_1}
\land \ldots \land \phi_{G_k,p_k}$.
\end{enumerate}
If $h=1$, then it is easy to construct
such a structure.  Given $i_1, \ldots, i_k$ such that $i_j
\in G_j$ (where the $i_j$ are not necessarily distinct)
we construct a structure $M$ with states $s, t_1,
\ldots, t_k$ (where $t_j = t_{j'}$ if $i_j = i_{j'}$)
such that $(M,t_j) \sat \band_{\{j': i_{j'} = i_j\}} \neg p_{j'} \land
\band_{\{j': i_{j'} \ne i_j\}} p_{j'}$, $(M,s)
\sat p_1 \land \ldots \land p_k$, and $\K_{i}$ is the least equivalence
relation that includes $(s,t_{j})$ if $i = i_j$.  It
is easy to check that $M$ has the required properties.  For the
inductive step, suppose that we are given $i_1, \ldots, i_k$ such that
$i_j
\in G_j$. Note that the first conjunct of $\phi_{G_j,p_j}$ has the
form $\neg E_{G_j'} \neg (\neg p_j \land \band_{k=1}^{m_j}
\phi_{G_{jk},p_{jk}})$.  By the induction hypothesis, we can find a
structure $M_j$ with state space $S_j$ and a state $t_j$  in  $S_j$ with
the properties above
such that $(M,t_j) \sat p_j \land
\band_{k=1}^{m_j}\phi_{G_{jk},p_{jk}}$. If
$i_j
= i_{j'}$, we can also assume without loss of generality that $M_j =
M_{j'}$ and $t_j
= t_{j'}$. Let $S$ consist
of $S_1 \union \ldots \union S_k$ together with a new state $s$.  We
define $M \in \MPrst$ so that each of the structures $M_j$ is embedded
in $M$ at
$(s,t_j)$ and the relation in $\K_{i_j}$ in $M$ is the least
equivalence relation that makes this true such that $(s,t_j) \in
\K_{i,j}$.  We leave it to the reader to check that we can define an
interpretation $\pi'$ with all the required properties.  Of course, the
fact that $\phi_{G,p}$ is satisfiable if $|G| > 0$ is now immediate.

{F}inally, define $\psi_G$ to be $\phi_{G,p} \land E_{G'}(q \land
(\neg p \rimp \phi_{G,q}))$, where we assume that the primitive
propositions that
appear in $\phi_{G,p}$ and $\phi_{G,q}$ are disjoint.

We claim that
$\psi_G$ is not satisfiable if $|G| \le 1$.  Clearly it is not
satisfiable if $|G| = 0$, since $\phi_{G,p}$ is not.  So suppose, by way
of contradiction, that $G = \{i\}$ and $(M,s) \sat \psi_G$ for some $M
\in \MPrt$. Then,
thanks to the properties of $\phi_{G,p}$ and $\phi_{G,q}$, we must have
$(M,s) \sat \neg K_i p \land K_i(q \land (\neg p \rimp \neg K_i q))$.
It is easy to see that this gives us a contradiction.  On the other
hand, if $|G| > 1$, we can construct a structure satisfying $\psi_G$ as
follows.  Suppose that $i, j \in G$ and $\phi_{G,p}$ is of the form
$$\neg E_{G'} \neg (\neg p \land \band_{j=k'+1}^k \phi_{G_j,p_j}) \land
E_{G_1} p \land \ldots \land E_{G_{k'}} p \land \band_{j=k'+1}^k
E_{G_j'} p_j.$$
We know that $|G_{k'}| = \cdots = |G_k| = 1$, so by
our previous argument, we can find a
structure $M' =(S', \ldots) \in \MPrst$ and states $s',t' \in S'$ such
that $(M,s') \sat \phi_{G,q} \land \band_{j=k'+1}^k \phi_{G_j,p_j}$,
$(s',t') \in \K_j$, $(M',t') \sat \neg q$, and $\K_i(s') = \{s'\}$.
Since $p$ does not appear in $\phi_{G,q}$, we can assume
without loss of generality that $(M,s) \sat \neg p$.  Now let $M \in
\MPrst$ be a structure whose state space
is $S' \union \{s\}$, where $s$ is a fresh
state not in $S'$, such that $M'$ is embedded in $M$ at $(s,s')$,
$(s,s') \in \K_i$, $(M,s) \sat p \land q$, and $\K_{i'}(s) = \{s\}$ for
$i' \ne i$.  It is easy to see that $(M,s) \sat \psi_G$.

{F}or part (c),
we construct formulas
$\phi_{m,G,p}$ such that
\begin{itemize}
\item if $(M,s) \sat \phi_{m,G,p}$ for $M \in \MPelt$ (and hence also
for
$M \in \MPrst$), then there exist $m+1$ distinct agents $i_1, \ldots,
i_{m+1} \in G$ such that $(M,s) \sat \neg K_{i_j} \neg p$, $j = 1,
\ldots, m+1$;
\item $|\phi_{m,G,p}|= O(l(G) + m)$;
\item if $|G| > m$, then $\phi_{m,G,p}$ is satisfiable in
$\MPrst$ (and hence in
$\MPelt$).
\end{itemize}

We first define an auxiliary family of formulas.
If $G', G_1, \ldots, G_k
\subseteq \G_\A$,
let $\psi_{m,G',G_1, \ldots,
G_k,p}$ be the formula
$$\begin{array}{ll}
E_{G_1} q_0 \land \ldots \land E_{G_k} q_0 \land\\
\neg E_{G'}\neg (p_0 \land p_1 \land q_1  \land E_{G'}(p_0
\rimp p_1 \land q_1)) \land\\
\ \ \ \ldots \land \neg E_{G'} \neg (p_0 \land p_{m+1} \land q_{m+1}
\land E_{G'}(p_0 \rimp p_{m+1} \land q_{m+1})) \land\\
E_{G'} ((p_0 \rimp (p \land \neg q_0)) \land
(q_1 \dimp \neg p_2 \land q_2) \land (q_2 \dimp \neg
p_3
\land q_3) \land \ldots \land (q_{m+1} \dimp {\em true})),
\end{array}$$
where $p_0, \ldots, p_{m+1}, q_0, \ldots, q_{m+1}$ are fresh primitive
propositions distinct from $p$.
%
Observe that $|\psi_{m,G',G_1,\ldots,G_k,p}|$ is $O(k + m)$.
It is easy to check that the last clause forces $q_i$,
for $1 \leq i \leq m$,
to be equivalent
to $\neg p_{i+1} \land \ldots \land \neg p_{m+1}$. at least in the
worlds $G'$-reachable in one step.
Thus, in these worlds, the formulas
$p_i \land q_i$, $i=1, \ldots, m+1$, are mutually exclusive.  Clearly if
$(M,s) \sat
\psi_{m,G',G_1,\ldots,G_k,p}$
for $M \in \MPelt$, then
there must be agents $i_1,
\ldots, i_{m+1}$ in $G' - (G_1 \union \ldots \union G_k)$ such that $(M,s)
\sat \neg K_{i_j} \neg (p_0 \land p_j \land q_j \land
E_{G'}(p_0 \rimp p_j \land q_j))$.
(Note that we cannot have $i_j \in
G' \inter (G_1 \union \ldots \union G_k)$ since $(M,s) \sat E_{G_j}q_0
\land E_{G'}(p_0 \rimp \neg q_0)$).   Thus, there must
exist states $t_j$, $j= 1,\ldots, m+1$ such that $(s,t_j) \in \K_{i_j}$
and $(M,t_j) \sat p_0 \land p_j \land q_j
\land E_{G'} (p_0 \rimp p_j
\land q_j)$.  To see that these
agents $i_j$ must be distinct, suppose that $i_j = i_{j'}$ for $j < j'$.
By the Euclidean property, we have $(t_j,t_{j'}) \in \K_{i_j}$.  Since
$(M,t_j) \sat E_{G'} (p_0 \rimp p_j \land q_j)$,
we must have $(M,t_{j'}) \sat p_j \land q_j$.  But since $(M,t_{j'})
\sat q_j \dimp (\neg p_j \land \ldots \land \neg p_{m+1})$, this is
inconsistent with the fact that $(M,t_{j'}) \sat p_{j'}$.
Since $(M,s) \sat E_{G'} (p_0 \rimp p)$, it follows that $(M,s) \sat
\neg K_{i_j} \neg p$ for $j = 1, \ldots, m+1$.
Conversely, it is easy to see that if $|G' - (G_1 \union \ldots \union
G_k)| > m$ then
$\psi_{m,G',G_1,\ldots,G_k,p}$
is
satisfiable in $\MPrst$.  We leave
the
details to the reader.

We now construct the formulas $\phi_{m,G,p}$
by induction on the least $h$
such that $G =  G' - \union \H$ and $\H \subseteq
(\G_\A)^m_h$.  If $\H = \{G_1, \ldots, G_k\} \subseteq
(\G_\A)^m_0 = \G_A$, then we take $\phi_{m,G,p} =
\psi_{m,G',G_1,\ldots,G_k,p}$.
Now suppose that $\H \subseteq (\G_\A)^m_h$ for $h >1$.
Without loss of generality, we can assume that $\H = \{G_1,
\ldots, G_{k'}, G_{k'+1}, \ldots, G_k\}$, where
$G_1, \ldots, G_{k'} \in
\G_\A$ and, for $j = k'+1, \ldots, k$,
$G_j \in (\G_\A)^m_{h-1}$ is of the
form $G_j' - \union \H_j$ with $G'_j \in \G_\A$, $\H_j \subseteq
(\G_\A)^m_{h-1}$, and $|G_j| \le m$.  Suppose that $|G_j| = m_j$.
By induction, for $j = k'+1, \ldots,k$,
we can construct formulas $\phi_{m_j-1,G_j,p}$
such that if $(M,s)\sat \phi_{m_j-1,G_j,p}$, then for each
agent $i \in G_j$,
we have $(M,s) \sat \neg K_i \neg p$ and
the
formula $\psi_{m,G',G_1,
\ldots, G_{k'},p}$.  Without loss of generality, we
can
assume that, other than $p$, the sets of primitive propositions
mentioned in the formulas $\phi_{m_j-1,G_j, p}$ are disjoint, and these
sets are all disjoint from the set of primitive propositions in
$\psi_{m,G',G_1,\ldots, G_{k'},p}$.
Let $\phi_{m,G,p}$ be the formula
$$
\psi_{m,G',G_1,\ldots,G_{k'},p'} \land
\band_{j=k'+1}^m \phi_{m_j-1,G_j, p} \land E_{G'}(p' \rimp E_{G'} \neg
p).
$$
The argument that this formula has the required properties is almost
identical to that for $\psi_{m,G',G_1, \ldots, G_k,p}$; we leave details
to the reader.

{F}inally, for part (d), consider the formula $\phi_d$ defined as
$$E_{G_1}p_1 \land
\ldots \land E_{G_{k-1}} p_{k-1} \land E_{G_k} (\neg p_1 \lor \ldots
\lor
\neg p_{k-1}).$$
We leave it to the reader to check that $\phi_d$ is satisfiable
in $\MPelt$ iff $G_1 \inter \ldots \inter G_k = \emptyset$.
\eprf


We already saw that for axiom E2 to be recursive, we need to be able to
decide whether
$|G_0 - (G_1 \union \ldots \union G_k)| \ge 1$ (or, equivalently,
whether $G_0 \subseteq G_1 \union \ldots \union G_k$) for $G_0, \ldots,
G_k \in
\G_\A$.
Proposition~\ref{reduction} shows that if there is no recursive
algorithm for answering such questions, the satisfiability problem for
the logic (even without $C_G$ operators) is also not decidable.
{F}or simplicity here, we assume we have oracles that can answer the
questions that we need to answer (according to
Proposition~\ref{reduction}) in unit time; we consider the
complexity of querying the oracle in more detail in
Section~\ref{oracle}.
More precisely, let $O_m$ be an oracle that, for a set $G \in \hG_\A^m$,
tells us whether $|G| > k$, for any $k < m$.  (Thus, queries to oracle
$O_m$ have the form $(G,k)$.)
Let $O'$ be an oracle that tells us whether
$G_1 \inter \ldots \inter G_k = \emptyset$, for $G_1, \ldots, G_k \in
\G_\A$.


\thm\label{complexity} There is a constant $c > 0$
(independent of $\A$)
and an algorithm that,
given as input a formula $\phi \in \LGC$,
decides if $\phi$
is satisfiable in
$\MP$ (\respc
$\MPr$, $\MPrt$, $\MPrst$, $\MPelt$) and runs in time
$2^{c|\phi|}$
given oracle $O_0$ (\respc $O_0$, $O_1$, $O_{|\phi|}$,
both $O_{|\phi|}$
and
$O'$), where queries to the oracle take unit time.   Moreover, if $\G$
contains
a subset with at least two elements, then there exists a constant $d > 0$
(independent of $\A$)
such that every algorithm for deciding the satisfiability of formulas
in $\MP$ (\respc $\MPr$, $\MPrt$, $\MPrst$, $\MPelt$) runs in time at
least $2^{d|\phi|}$, even given access to oracle $O_0$
(\respc $O_0$, $O_1$, $O_{|\phi|}$, both $O_{|\phi|}$ and $O'$),
for infinitely many formulas $\phi$.
\ethm


Before proving
Theorems~\ref{complete} and~\ref{complexity}, we prove a somewhat
simpler theorem that
allows us to both
explain intuitively why the
results are
true and
point out some of the
difficulties in proving them.

\pro\label{Shore} If there is an oracle that
decides if $G = \emptyset$
for each Boolean combination $G$ of elements in $\G_\A$,
then,
for every
formula $\phi \in \LGC$, we can effectively find a formula $\phi^\sigma$
in a language $\LGp$, where $\G'$ consists of all subsets
of
a set
$\A'$ of
at most $2^{|\phi|}$ agents, such that $|\phi^\sigma| = |\phi|$ and
$\phi$
is satisfiable in $\MP$ iff $\phi^\sigma$ is satisfiable in $\M_{\A'}$.
\epro
\prf
Given $\phi$, let $\G_\phi$ be the set of subsets
$G$ of agents such that $E_G$ or $C_G$ appears in $\phi$.
(Recall that we are identifying $K_i$ with $E_{\{i\}}$,
so that $\{i\} \in \G_\phi$ if $K_i$ appears in $\phi$.)
Note that $|\G_\phi| \le |\phi|$.

Suppose that $\G = \{G_1, \ldots, G_N\}$.
An
{\em atom over
$\G$\/} is a nonempty set of the form $G_1' \inter \ldots \inter
G_N'$, where $G_i' = G_i$ or $G_i' = \overline{G_i}$.
Clearly there are at most $2^N$ atoms over $\G$.
Let $\A'$ consist of the
nonempty atoms over $\G_\phi$.  Note that $|\A'| \le 2^{|\phi|}$.
Define $\sigma: \A
\rightarrow \A'$ by taking $\sigma(i)$ to be the unique atom over
$\G_\phi$ containing $i$.  We extend $\sigma$ to a map from $2^{\A}
\rightarrow 2^{\A'}$ by taking $\sigma(G) = \{\sigma(i): i \in G\}$
($= \{H\in \G_\phi: H \subseteq G\}$).
Translate $\phi$ to
$\phi^\sigma$ by replacing all occurrences of
$E_G$ and $C_G$ in
$\phi$ by
$E_{\sigma(G)}$, and
$C_{\sigma(G)}$, respectively.
Clearly $|\phi| = |\phi^\sigma|$.  (Note that it is important here that
we take the length of $E_G$ and $C_G$ to be
independent of $G$.)

If $\phi $ is satisfiable, let $(M,s)$ witness that fact.  Convert $M$
into a structure $M^\sigma$ over $\A'$ with the same state space by
setting $(s,t)\in \K_A$ iff
$(s,t)\in \union_{j \in A}\K_{j}$
for each $A \in \A'$.
An easy induction shows that for every formula $\psi $
with
sets
(of agents)
chosen from
$\G_\phi$,
we have $(M,s)\models \psi $ if and only if $%
(M^\sigma,s)\models \psi^\sigma$. The only point that needs any comment
is that
$E_{G}$ (and so also $C_{G})$ has the same meaning in $M$ (in terms of
reachability)
as $E_{\sigma(G)}$ ($C_{\sigma(G)}$) in $M^{\sigma}$,
by the definition
of $\sigma(G)$ and the $\K_{A}$ relations. Thus $(M^{\sigma},s)\models
\phi^{\sigma}$ as required.

{F}or the other direction, suppose that $(M',s)\models \phi^\sigma$ for
some
structure $M'$ over $\A'$.   We define a
structure $M$ over $\A$ by defining $\K_i = \K_{\sigma(i)}$.
Again an easy induction shows that for every formula $\psi $ with
sets chosen from
$\G_\phi$,
$(M^{\prime },s)\models \psi $ if and only if $(M,s)\models
\psi^{\sigma}$. Once again, the only point to notice is that
$E_{G}$
(and so also $C_{G}$) has the same meaning in $M^{\prime }$ (in
terms
of reachability) as $E_{\sigma(G)}$ ($C_{\sigma(G)}$) in $M$ by the
definition of $\sigma(G)$ and the relations $\K_{j}$. Thus $(M,s)\models
\phi $ as required. \eprf

\cor\label{Shore1} Given an oracle that
decides, for each Boolean combination $G$ of elements in
$\G_\A$,
whether
$G = \emptyset$, there is
a constant $c > 0$ (independent of $\A$) and
an algorithm that,
given as input a formula $\phi \in \LGC$,
decides if $\phi \in \LGC$ is satisfiable in $\MP$ and runs in time
$2^{c2^{|\phi|}}$. \ecor

\prf Clearly, to check if $\phi$ is satisfiable, it suffices to check if
$\phi^\sigma$ is satisfiable.  In \cite{HM2}, there is an exponential
time
algorithm for checking satisfiability.  However, this algorithm presumes
that the set of agents is fixed.  A close look at the algorithm actually
shows that it runs in time $2^{cm|\phi|}$, where $m$ is the number of
agents.  In our translation, the set of agents is exponential in
$|\phi|$,
giving us a double-exponential
time
algorithm.
\eprf

\cor\label{Shore2} If $\G$ is closed under
intersection
and complementation,
then $\Kaxc$ is a sound and complete axiomatization for the language
$\LGC$ with respect to $\MP$. \ecor

\prf  Soundness is straightforward, so we focus on completeness.
Suppose that $\phi$ is valid.  By Proposition~\ref{Shore}, so is
$\phi^\sigma$. Since $\A'$ is finite,
$\Kaxcp$ is a complete axiomatization for $\LGp$ with respect to
$\M_{\A'}$.  Thus, $\Kaxcp \vdash \phi^\sigma$.   We
can translate this proof step by step to a proof of $\phi$ in $\Kaxc$.
We simply replace every formula $\psi$ that appears in the proof of
$\phi^{\sigma}$ by $\psi^{\tau}$, where $\psi^{\tau}$ is
obtained by replacing each occurrence
of
$K_A$ in $\psi$
by $E_{A}$ unless $A = \{i\}$ is a singleton, in which case
we replace $K_A$ by $K_i$, and replacing each occurrence of
$E_G$, and $C_G$ in $\psi$ by
$E_{\union G}$, and $C_{\union G}$, respectively.
Since we have assumed $\G$ is closed under complementation and
intersection, it is also closed under union, and hence $\psi^\tau$ is
a formula in $\LGC$.

It is easy to check that the translated proof is still a proof
over the language $\LGC$:
Tautologies become tautologies as $(\phi \vee \psi )^{\tau}=\phi
^{\tau}\vee
\psi^{\tau}$ and similarly for negations.
Instances of MP in the proof of $\phi^\sigma$ become instances of
MP in the proof of $\phi$
because $(\phi \rightarrow \psi )^{\tau}=\phi^{\tau}\rightarrow
\psi^{\tau}$.
Instances of KGen in the proof of $\phi^\sigma$ become instances of EGen
or KGen in the proof of $\phi$; similarly, instances of K1 are
converted to
instances of K1 or E1.
It is easy to see
that instances of E1, E2,
E3, EGen, C1, and RC1 are converted to legitimate instances of the same
axiom.
\eprf

While Corollaries~\ref{Shore1} and~\ref{Shore2} are close to our desired
theorems, they also make clear the difficulties we need to overcome in
order to prove Theorems~\ref{complete} and~\ref{complexity}.
Specifically,
\begin{itemize}
\item we need to cut the complexity down from double-exponential
to single exponential;
\item we need to prove completeness without assuming that $\G$ is
closed under complementation and intersection;
\item we want to use an oracle that tests only whether a set of
the form $G_0 - (G_1 \union \ldots \union G_k)$ is nonempty, rather
than one that applies to arbitrary Boolean combinations;
\item we want to extend these results to the case that the $\K_i$
relations satisfy properties like transitivity.
\end{itemize}
With regard to the last point, while in general it is relatively
straightforward to extend completeness and complexity results to deal
with relations that have properties like transitivity, it is not so
straightforward in this case.  For example, even if $M \in \MPrt$, the
relations in the structure $M^\sigma$ constructed in
Proposition~\ref{Shore} are not necessarily transitive.  As shown in
Proposition~\ref{reduction},
we need a different oracle to deal with transitivity.

\section{Proving the Main Results}\label{proofs}
In this section, we prove
Theorems~\ref{complete} and~\ref{complexity}.
The structure of the proof is similar to that of
Corollaries~\ref{Shore1} and~\ref{Shore2}; we describe step by step the
modifications required to deal with the problems raised in the previous
section.
It is convenient to split the proof into four cases,
depending on the class of structures considered.

\subsection{The Proof for $\MP$ and $\M_\A^r$}

In Proposition~\ref{Shore} we showed that we
could translate a formula $\phi$ to a formula $\phi^\sigma$ such that
$\phi$
was satisfiable in $\MP$ iff $\phi^\sigma$ was satisfiable in
$\M_{\A'}$,
where $\A'$
consisted
of the atoms over $\G_\phi$.  Our goal is to
maintain the translation idea, but use as our target set of agents
a set whose elements we can determine with the oracles at our disposal
(for testing the
nonemptiness of certain set differences).  As a first step, we
try
to abstract the key ingredients of Proposition~\ref{Shore}.  Suppose
that we have a set $\A'$ of agents and a
partial map
$\sigma: \A \rightarrow \A'
$.
Again, we can extend $\sigma$ to a map from $2^\A$ to
$2^{\A'}$: $\sigma(G) = \{\sigma(i): i \in \G
\}$.  Given a formula $\phi$, let $\phi^\sigma$ be the
formula
that results by replacing all the occurrences of $G$ in $\phi$ by
$\sigma(G)$.
In Proposition~\ref{Shore}, $\A'$ is the set of
atoms over $\G_\phi$ and $\sigma(i)$ is the unique atom containing
$i$.  We were able to show
that,
for that choice of $\A'$ and $\sigma$,
the formulas $\phi$ and $\phi^\sigma$ were
equisatisfiable.
What does
it take to obtain such a result in general?
The following result shows that we need to be able to
find a mapping $\tau: \A' \rightarrow 2^\A - \{\emptyset\}$ with one key
property.

\pro\label{trans} Given a formula $\phi$ and a
partial map
$\sigma: \A
\rightarrow \A'$ such that $\sigma(G) \ne \emptyset$ for all $G \in
\G_\phi$, suppose that there is a mapping $\tau: \A' \rightarrow
2^\A -
\{\emptyset\}$
such that for all $G \in \G_\phi$, we have
$\union \{\tau(A): A \in \sigma(G)\} = G$.
Then $\phi$ is satisfiable in $\MP$ (\respc $\MPr$) iff $\phi^\sigma$
is satisfiable in $\M_{\A'}$ (\respc $\M_{\A'}^r$).
\epro

\prf
Given $\phi$ and $\sigma$, suppose there exists a mapping $\tau$
with the property above.  We show that $\phi$ and $\phi^{\sigma}$ are
equisatisfiable.

{F}irst suppose that $(M,s)\models\phi$, where $M \in \MP$. We convert
$M = (S,\pi, \{\K_i: i \in \A\})$
into a structure
$M' = (S,\pi, \{\K_A: A \in \A'\})$
by defining $\K_A =
\union \{\K_i:i\in\tau(A)\}$.  Notice that the
assumed
property of
$\tau$ implies that for all $G \in \G_\phi$, we have
$$\union_{A \in \sigma(G)} \K_A = \union_{A \in \sigma(G)} \union_{i \in
\tau(A)} \K_i = \union_{i \in G} \K_i.$$
An easy induction on
the structure of $\psi$ now shows that $(M,t)\models\psi$ if and
only if $(M',t)\models\psi^{\sigma}$ for all $t \in S$ and all formulas
$\psi
\in {\cal L}_{\G_\phi}^C$.
Also note that if $M \in \MPr$, then $M' \in \M_{\A'}^r$ (since the
union of reflexive relations is reflexive).

{F}or the opposite direction, suppose $(M',s)\models\phi^{\sigma}$ for
some
$M' = (S,\pi, \{\K_A: A \in \A'\}) \in \M_{\A'}$.  Define $M =
(S,\pi,\{\K_i: i \in \A\}) \in \MP$ by setting
$\K_i=\K_{\sigma(i)}$
if $\sigma(i)$ is defined and the empty relation otherwise.
Note that for all $G \in \G_\phi$ we have
$$\union_{i \in G}\K_i = \union_{i \in G} \K_{\sigma(i)}  = \union_{A
\in \sigma(G)} \K_A.$$
Again, an easy induction on
the structure of $\psi$
shows that
$(M,t)\models\psi$ if and
only if $(M',t)\models\psi^{\sigma}$ for all $t \in S$ and all formulas
$\psi \in {\cal L}_{\G_\phi}^C$.

If $M' \in \MPr$, we modify the construction slightly by taking
$\K_i = \{(t,t): t\in S\}$ if $\sigma(i)$
is undefined.
Since
$\sigma(G)
\ne \emptyset$ for $G \in \G_\phi$, it is easy to check that we still
have $\union_{i\in G} \K_i = \union_{i \in G} \K_{\sigma(i)}$, so the
modified construction works for the reflexive case.
\eprf

{F}or the mapping $\sigma$ of Proposition~\ref{Shore} we
can take $\tau$ to be the identity, but this requires an oracle for
nonemptiness of atoms.
We now show how to choose $\A'$ and define maps $\sigma$ and $\tau$ in a
way that requires
only information about whether sets of the form $G_0 - (G_1 \union
\ldots \union G_k)$
are empty.

\dfn\label{Gmax}
Given a set $\G$ of sets of agents and
$G \in \G$,
a set $\H \subseteq \G$ is a {\em $G$-maximal\/} subset of $\G$
if $G - \union \H \ne \emptyset$ and
$G - ((\union \H)
\union G')
=\emptyset$ for all
$G'
\in \G -\H$.
Let $\R(\G)=\{(G,\H):G\in\G, \H \mbox{ is a $G$-maximal subset of
$\G$}\}$.~\edfn

Note that we can check whether $\H$ is a $G$-maximal subset of $\G$ by
doing at most
$|\G|$ tests of the form $(G - \union \H') = \emptyset$, and we can find
all pairs $(G,\H)$ in $\Rmax$ by doing at most $|\G|2^{|\G|-1}$ such
tests.

The following lemma gives some technical properties of $\R(\G)$ that
will be used frequently.

\lem\label{unique}
Suppose that $(G,\H) \in\R(\G)$ for some set $\G$ of
subsets of $\A$.
\begin{itemize}
\item[(a)] $G - \union \H$ is an atom over $\G$ and, in fact,
$G - \union \H = \inter(\G - \H) \inter
(\inter_{H \in \H} \overline{H})$.
\item[(b)] If $(G',\H) \in \R(\G)$, then
$(G-\bigcup\H)=(G'-\bigcup\H)$.
\item[(c)] If $(G',\H') \in \R(\G)$
and
$\H \ne \H'$, then
$(G-\bigcup \H)\cap(G'-\bigcup\H')=\emptyset$.
\end{itemize}
\elem

\prf
{F}or part (a), first observe that
since $\H$ is a $G$-maximal
subset of $\G$, for $H \notin \H$, we have $G - \union
(\H \union \{H\}) = \emptyset$; \ie $G - \union \H \subseteq H$.
Thus, if $H \notin \H$, we have $G - \union \H = (G \inter H) - \union
\H$.  Thus, $G -\union \H = G \inter
(\inter_{H \in \H} \overline{H}) = \inter(\G - \H) \inter
(\inter_{H \in \H} \overline{H})$, as desired.  By definition, $G - \union
\H$ is an atom over $\G$.

Part (b) is immediate from part (a), since it is clear
that
$G - \union \H$
is independent of $G$ and depends only on $\H$.


{F}or part (c), suppose that $\H \ne \H'$. Without loss
generality, there is some $H \in \H - \H'$.  It follows
immediately from part
(a) that $G - \union \H$ and $G' - \union \H'$ are distinct atoms (hence
disjoint), since $G - \union \H \subseteq \overline{H}$ and $G' \union \H'
\subseteq H$.
\eprf


If $(G,\H) \in \R(\G)$, let $A_{\H}^\G$ denote the atom associated with
$\H$ defined in Lemma~\ref{unique}(a). It is independent of $G$ by
Lemma~\ref{unique}(b). We omit $\G$,
writing simply $A_\H$,
when it is clear from the context which set $\G$
we have in mind.

We now show how to define a translation satisfying the hypotheses of
Proposition \ref{trans} using the elements of $\Rmax$ identified
according to the second coordinate alone.

Given a formula $\phi$, let $\A^\phi=\{\H: \exists
G[(G,\H)\in\Rmax]\}$.  Define $\sigmak: \A \rightarrow \A^\phi$
by setting
$\sigmak(i) = \H$ if
$i \in A_\H$ (as defined after Lemma~\ref{unique})
and undefined otherwise.
As before, we extend $\sigmak$ to $2^\A$
by
defining $\sigmak(G) =
\{\sigmak(i): i \in G
\}$.

\lem\label{K}
Define $\tau: \A^\phi \rightarrow 2^\A$ by setting $\tau(\H) = \inter
(\G_\phi - \H)$. Then
\begin{itemize}
\item[(a)] $\sigmak(G) =
\{\H \in \A^\phi: \exists G' \in \G_\phi((G',\H) \in \Rmax), \, G
\notin \H\}$,
\item[(b)] $\sigmak(G) \ne \emptyset$ for $G \in \G_\phi$,
\item[(c)] $\tau(\H) \ne \emptyset$ for $\H
\in \A^\phi$,
\item[(d)] $\union \{\tau(\H): \H \in \sigmak(G)\} = G$.
\end{itemize}
\elem

\prf  For part (a),
first suppose that $G \notin \H$
and $(G',\H) \in \R(\G_\phi)$ for some $G' \in \G_\phi$.
Then by Lemma~\ref{unique}(a), it
follows that $A_\H \subseteq G$.  Since $A_\H \ne \emptyset$, there is
some $i \in A_\H$.  Since $i \in G$ and $\sigmak(i) = \H$, it follows
that
$\H
\in \sigmak(G)$.  For the opposite inclusion, suppose that
$\H
\in \sigmak(G)$.  Then
$\H
= \sigmak(i)$ for some $i \in G
\inter A_\H$.  Since $G \inter A_\H \ne\emptyset$, it follows from the
definition of $A_\H$ that $G \notin \H$.

{F}or part (b), given $G$, note that there must be some $G$-maximal
subset $\H$.
Thus, $(G,\H) \in \R(\G_\phi)$.
Since $G - \union \H \ne \emptyset$, we must have $G
\notin \H$.  By part (a), $\H \in \sigmak(G)$, so $\sigmak(G) \ne
\emptyset$.

{F}or part (c), suppose that $\H \in
\A^\phi$.
  Then there exists some $G$ such
that $(G,\H) \in \Rmax$, and hence $G - \union \H \ne \emptyset$.  It
suffices to show that $\inter (\G_\phi - \H) \supseteq G - \union \H$.
Since $G - ((\union \H) \union G') = \emptyset$ for all $G'
\in \G_\phi - \H$, it follows that
$G - \union \H \subseteq G'$ for each $G' \in \G_\phi - \H$.
Thus, $\inter (\G_\phi - \H) \supseteq G - \union \H$.

{F}or part (d), we first show that
$\union \{\tau(\H): \H \in \sigmak(G)\} \subseteq G$.  Note that if $\H
\in \sigmak(G)$, then by part (a),
$G \in \G_\phi - \H$.  Thus, $\tau(\H) =
\inter (\G_\phi -\H) \subseteq G$.
{F}or the oppposite containment, suppose that $i \in G$.  Let $\H^i =
\{G' \in \G_\phi: i \notin G'\}$.  Since $i \in G - \union \H^i$,
there must be a $G$-maximal subset $\H$ of $\G_\phi$ containing $\H^i$.
By part (a), we have $\H \in \sigmak(G)$.
Moreover,
since $\H^i \subseteq \H$, for all $H' \in \G_\phi - \H$, we have $i \in
H'$.  Thus, $i \in \inter (\G_\phi - \H)$.  It follows that $ i \in
\union_{\H \in \sigmak(G)} \inter (\G_\phi - \H)$, as desired. \eprf

Since $|\A| \le 2^{|\phi|}$, we have now reduced satisfiability with
infinitely many agents to satisfiability with finitely many agents, at
least for $\MP$ and $\MPr$, using only tests that we know we need
to be able to perform in any case.   We next must deal with the problem
we observed in the proof of Corollary~\ref{Shore1}, that
is,
there may be
exponentially
many
agents in the subgroups mentioned in $\phi^{\sigmak}$.  This is
done in the following result.  In this result, we assume that the
complexity of checking whether $i \in G$ is no worse than linear in
$|\A|$.  While we do not assume this in general, it is true for the
$\A'$ and sets $G$ that arise in the translation of
Proposition~\ref{trans}, which suffices for our application of
the result to the proof of Theorem~\ref{complexity}.

\thm\label{dec}
If $\A$ is finite and there is an algorithm for deciding if $i \in G$
for $G \in \G$ that runs in time linear in $|\A|$, then
there is
a constant $c > 0$ (independent of $\A$)
and an algorithm that,
given as input a formula $\phi \in \LGC$,
decides if $\phi$ is satisfiable in
$\MP$ (\respc $\MPr$)
and runs in time $O(|\A|2^{c|\phi|})$.
\ethm

\prf We first present an algorithm that
decides if $\phi$ is satisfiable in
$\MP$; we then show how to modify it to deal with $\MPr$.  The
algorithm is just
a slight modification of standard decision procedures \cite{FHMV,HM2}.
(Far more serious modifications are needed to prove
the analogous result for the $\MPrt$, $\MPrst$, and $\MPelt$; see
Theorems~\ref{decrt}, \ref{decrst}, and~\ref{decelt}.)

Let $\Sub(\phi)$ be the set of subformulas of $\phi$ together with
$E_{G}(\psi \land C_G \psi)$ and $\psi \land C_{G}\psi$ for each
subformula $C_{G}\psi$ of
$\phi$. $\Sub^{+}(\phi)$ consists of the formulas in $\Sub(\phi)$ and
their negations.  An easy induction
on $|\phi|$ shows that $|\Sub(\phi)| \le |\phi|$, so $|Sub^+(\phi)| \le
2|\phi|$.  (Here we need to use the fact that we take the length of
$C_G$ to be 3.)

Let $S^1$ consist of all
subsets $s$ of $\Sub^+(\phi)$ that are
{\em maximally consistent\/} in that
(a) for each formula $\psi
\in \Sub(\phi)$, either $\psi \in s$ or $\neg \psi
\in s$,
(b)
they are
propositionally consistent
(for example, we cannot have
all of $\psi \land \psi'$, $\neg \psi$, and $\neg
\psi'$ in $s$), and
(c) they contain $E_G (\psi \land C_G \psi)$
iff
they contain $C_G \psi$.
Note that there are at most $2^{|\phi|}$ sets in $S^1(\phi)$.

{F}or $s \in S^{1}$ and $G \in
\G_\A$, we define $s/E_G=\{\psi:
E_G\psi\in s\}$ (again, we identify $K_i$ with $E_{\{i\}}$).
Define $s/\oKi = \union_{i \in G} (s/E_G)$.
Define
a
binary relation $\K_i$ on $S^1$ for each $i \in \A$ by taking
$(s,t) \in\K_i$ iff $s/\oKi \subseteq t$.
We now define a
sequence $S^{j}$ of subsets of $S^{1}$.
Suppose that we have defined $S^1, \ldots, S^j$.  $S^{j+1}$ consists of
all states in $S^j$ that {\em seem consistent}, in that
the following two
conditions hold:
\begin{enumerate}
\item If $\lnot E_{G}\psi\in s$, then there is some $t \in S^j$
such that $(s,t) \in \union_{i \in G}\K_i$ and $\neg \psi \in t$.
\item If $\lnot C_{G}\psi\in s$,
then
there is some $t \in S^j$ such that
$t$ is $G$-reachable from $s$ in $S^j$ and $\neg \psi \in t$.
\end{enumerate}
If $S^j \ne S^{j+1}$
then we continue the construction.  Otherwise the construction
terminates; in this case, the algorithm returns
``$\phi$ is satisfiable'' if $\phi \in s$ for some state
$s \in S^{j+1}$ and returns ``$\phi$ is unsatisfiable'' otherwise.

Since $S^{j} \supseteq S^{j+1}$, $S^{1}$ has at most
$2^{|\phi|}$ elements, and there are $|\A|$
relations,
it is easy to see that the whole procedure can be carried out
in time $O(|\A|2^{c|\phi|})$ for some $c > 0$.

It remains to show that the algorithm is correct.
{F}irst suppose that
$\phi$ is satisfiable.  In that case, $(M,s_0)\sat \phi$ for some
structure $M = (S,\pi, \{\K_i': i \in \A\})
\in \MP$.  We can associate with each state $s \in S$ the state $s^*$ in
$S^1$ consisting of all the formulas $\psi \in \Sub(\phi)$ such that
$(M,s) \sat \psi$.  It is easy to see that if $(s,t) \in \K_i'$ then
$(s^*,t^*) \in \K_i$.   A straightforward
induction shows that the states $s^*$ for $s \in S$ always seem
consistent, and thus are in $S^j$ for all $j$. Moreover, $\phi \in
s_0^*$.  Thus, the algorithm declares that $\phi$ is satisfiable, as
desired.

Conversely, suppose that the algorithm declares that $\phi$ is
satisfiable.
We construct a structure $M = (S,\pi, \{\K_i': i \in \A\})$ over $\A$
and $\Phi$ in which
$\phi$ is satisfied as follows.  Let $j$ be the stage at which the
algorithm terminates.  Let $S = S^j$.
Define $\pi$ so that $\pi(s)(p) = {\bf true}$ iff $p
\in s$, for $s \in S$ and $p \in \Phi$.  For each $i \in \A$,
we take $\K_i'$  to be the restriction of $\K_i$ to $S^j$.
A straightforward induction on the structure of formulas shows that
for all formulas $\psi \in \Sub(\phi)$ and states $s \in S$, we have
$(M,s) \sat\psi$ iff $\psi \in s$.
(The cases for $E_G \psi$ and $C_G \psi$ use the appropriate clauses of
the definition of seeming inconsistent and the choice of $j$.)
Since $\phi \in s$ for some $s^* \in
S$, it follows that $(M,s^*) \sat \phi$, so $\phi$ is satisfiable.

To deal with $\MPr$, the only change necessary is that in going
from $S^1$ to $S^2$ in the construction, we also eliminate $s \in S^1$ if
$(s,s) \notin \K_i$ for some $i \in \A$.  This guarantees that the
$\K_i$ relations are reflexive.  The remainder of the
proof goes through unchanged.
\eprf


\bigskip

\noindent
{\bf Proof of Theorem~\ref{complexity} for $\MP$ and $\MPr$:}
The deterministic exponential time lower bound in
Theorem~\ref{complexity} follows from the lower bound in the case
where $\A$ is finite, which is proved in \cite[Theorem~6.19]{HM2} using
techniques developed by Fischer and Ladner \citeyear{FL} for PDL.
The sets $G$ that arise in the lower bound proof have cardinality 2, so
oracles are of no help here.

{F}or the upper bound, suppose that we are given a formula $\phi$.
We first compute the set $\Rmax$.  This can be done with
at most $|\phi|2^{|\phi|}$
calls to oracle $O_0$,
since $|\G_\phi| \le |\phi|$ and we
need only check,
for each $G \in \G_\phi$ and $\H \subseteq \G_\phi$, whether $G - \H =
\emptyset$.

Consider the mapping $\sigmak$ of Lemma~\ref{K}.  By part (a)
of Lemma~\ref{K}, we can compute the formula $\phi^{\sigmak}$
using $\le |\phi|2^{|\phi|}$ calls to oracle $O_0$.  By
Proposition~\ref{trans} and
Lemma~\ref{K}, the formulas $\phi$ and $\phi^{\sigmak}$ are
equisatisfiable.  By Theorem~\ref{dec}, we can decide if $\phi^{\sigmak}$
is satisfiable in time $O(2^{c|\phi|})$ for some $c > 0$ (since
$|\phi^{\sigmak}| = |\phi|$ and the set $\A$ of agents that appear in
$\phi^{\sigmak}$ has
size
at most $2^{|\phi|}$).  \eprf

We now want to prove Theorem~\ref{complete} in the case of $\MP$ and
$\MPr$.  The idea is the same as that of Corollary~\ref{Shore2}.  If
$\phi$ is valid, then so is $\phi^{\sigmak}$.  We can then appeal to
completeness in the case of finitely many agents to get a proof of
$\phi^{\sigmak}$ that we can then ``pull back'' to a proof of $\phi$.
There is only one difficulty that we encounter
when
trying to put this idea
into practice.  Exactly how do we pull back the proof?  For example,
suppose that the proof of $\phi^{\sigmak}$ involves a formula $\psi$
with an
operator $K_\H$.  In general, there will be many agents $i \in \A$
such that $\sigmak(i) = \H$.  One option is to replace $K_\H$ by
$E_{\sigmak^{-1}(\H)}$, that is, replace $\H$ by all $i$ such that
$\sigmak(i) = \H$.  (This is what was done in the proof of
Corollary~\ref{Shore2}.)  The problem with this is that there is no
guarantee that the resulting set is in $\G$.  Alternatively, we could
replace $K_\H$ by $K_i$ for some $i$ such that $\sigmak(i) =\H$.  But if
so, which one?

We actually take the latter course here.  We solve the problem of which
$i$ to choose by showing that there is a proof of $\phi^{\sigmak}$ in
which the only modal operators that arise in any formula used
in the proof are modal operators that appear in $\phi^{\sigmak}$
(Lemma~\ref{proof}).  For these operators,
there is a canonical way to do the replacement
(Lemma~\ref{replace}).  While it may seem almost trivial that the only
operators that should be needed in the proof of $\phi^{\sigmak}$ are
ones that already appear in the formula, this is not the case for the
standard completeness proof \cite{FHMV,HM2}, since in the proof of the
validity of a formula of the form $E_G \psi$, the modal operators $K_i$
are used for $i \in G$, although these operators may not appear in
$\psi$.  It is important that we use the axioms E1 and E2 in doing the
proof, rather than the axiom E; otherwise the result would not hold.
Indeed, the result does not quite hold in the case of $\Taxc$; we need
to augment it with E5.

\lem\label{replace} The mapping $\sigmak$ (when viewed as a map with
domain $2^\A$) is injective on $\G_\phi$.
\elem

\prf Suppose that $G \ne G'$.  Without loss of generality, suppose that
$i
\in G - G'$.  Then there is a $G$-maximal set $\H$ that includes $G'$.
By Lemma~\ref{K}(a), we have $\H \in \sigmak(G)$.  Since $G' \in \H$, it
follows from Lemma~\ref{K}(a) that $\H \notin \sigmak(G')$.  Thus,
$\sigmak(G) \ne\sigmak(G')$.  \eprf

{F}or the next lemma, we write $AX \vdash_\phi \psi$ if there is a proof
of $\phi$ in AX that involves only modal operators that appear in
$\phi$. Let $\Taxcp$ consist of $\Taxc$
augmented with the axiom E5.
Although E5 follows from E1 and K2, using E5
allows us to be able to write proofs of $\phi$ that use only the modal
operators in $\phi$.

\lem\label{proof}  If $\A$ is finite and $\phi \in
\LGC$ is
valid with respect to $\MP$ (\respc $\MPr$), then $\Kaxc
\vdash_\phi \phi$ (\respc $\Taxcp \vdash_\phi \phi$). \elem

\prf We first consider the case of $\MP$.
Since $\phi$ is
valid, $\neg \phi$ is not satisfiable.  That means, when we apply the
construction in the proof of Theorem~\ref{dec} to $\neg \phi$, all the
sets containing $\neg \phi$ are eliminated.
{F}or each state $s \in S^1$, let $\phi_s$ be
the conjunction of all the formulas in $s$.

We prove the result by showing, by induction on $j$, that
\begin{equation}\label{eq0}
\mbox{if a state $s \in S^j$ does not seem
consistent, then $\phi_s$ is $\Kaxc$-inconsistent, \ie $\Kaxc
\vdash_\phi \neg \phi_s$.}
\end{equation}
To see that (\ref{eq0}) suffices to
prove the lemma,
note that
standard propositional reasoning
(i.e., using Prop and MP) shows that,
for any formula $\psi \in \Sub(\neg \phi)$,
$$
\Kaxc \vdash_\phi \psi
\dimp \lor_{\{s \in S^1: \psi \in s\}} \phi_s.
$$
(Here we need the observation that by C1 and RC1, nothing is lost by
our assumption that $C_G\psi \in s$ iff $E_G(\psi \land C_G\psi) \in
s$.) Negating both sides of $\dimp$, we get
\begin{equation}\label{eq-1}
\Kaxc \vdash_\phi \neg \psi
\dimp \land_{\{s \in S^1: \psi \in s\}} \neg \phi_s.
\end{equation}
Thus, if $\Kaxc \vdash_\phi \neg \phi_s$ for each set $s$ containing
$\neg
\phi$, it follows by standard propositional reasoning that $\Kaxc
\vdash_\phi \phi$, as desired.

While this general approach to proving completeness is quite standard,
we must take extra care because of our insistence on restricting to
symbols that appear in $\phi$, particularly when dealing
with the case when a state seems inconsistent due to a formula of the
form $\neg E_G \psi$
or $\neg C_G \psi$
not being satisfied.  This is where the axioms E1
and E2 come into play.
	
To prove (\ref{eq0}),
we first need a number of basic facts of epistemic logic and some
preliminary observations.
The basic facts (which are easily proved using Prop, E3 (or K1 when $G=\{i\}$),
MP, and EGen (or  KGen);
see \cite[p.~51, 94]{FHMV}) are that if $\psi$ and $\psi'$ involve only
modal operators in $\phi$,
then
\begin{equation}\label{basic3}
\Kaxc \vdash_\phi E_G(\psi \land \psi') \dimp E_G\psi \land E_G\psi'
\end{equation}
and
\begin{equation}\label{basic4}
\mbox{if $\Kaxc \vdash_\phi \psi \rimp \psi'$ then
$\Kaxc \vdash_\phi E_G \psi \rimp E_G \psi'$.}
\end{equation}


Assume by induction that for all $s \in S^1 - S^j$, we have $\Kaxc
\vdash_\phi \neg \phi_s$.
We now show that if $s \in S^j$ does not seem consistent
then $\Kaxc \vdash_\phi \neg \phi_s$, by considering in turn each of the
two ways $s$ may seem inconsistent.

{F}irst suppose that $s$ does not seem consistent because $\neg E_G \psi \in s$
and there is no state $t \in S^j$ such that $(s,t) \in \union_{i \in G}\K_i$
and $\neg \psi \in t$.  We show
that
\begin{equation}\label{eq5}
\Kaxc \vdash_\phi \phi_s \rimp E_G\psi.
\end{equation}
Since $\neg E_G \psi$ is a conjunct of $\phi_s$ (since $\neg E_G
\psi \in s$, by assumption),
(\ref{eq5})
shows that $\phi_s$ is
$\Kaxc$-inconsistent, as desired.

To prove (\ref{eq5}),
we first show that if $G \in \G_\phi$, then
\begin{equation}\label{eq2}
\mbox{if $(s,t) \notin \union_{i \in G}\K_i$,
then $\Kaxc \vdash_\phi \phi_s \rimp E_G \neg \phi_t$.}
\end{equation}

To prove~(\ref{eq2}), suppose that $(s,t) \notin \union_{i \in G} \K_i$.
{F}or each $i \in G$, there must be some $G^{i,t}
\in \G_\phi$ and formula $E_{G^{i,t}}
\theta$ such that $i \in G$, $E_{G^{i,t}} \theta \in s$ and $\neg \theta \in t$.
Since $E_{G^{i,t}} \theta \in s$ and $\neg \theta \in t$
it is immediate that
$\Kaxc \vdash_\phi \phi_s \rimp E_{G^{i,t}} \theta$ and
$\Kaxc \vdash_\phi \theta \rimp \neg \phi_t$.  Now applying
(\ref{basic4}) and propositional reasoning, we
get that $\Kaxc \vdash_\phi
\phi_s \rimp E_{G^{i,t}} \neg \phi_t$.  Since we can find such a $G^{i,t}$
for each $i \in G$, we have that $G \subseteq \union_{i \in G} G^{i,t}$.
Since $G$ is finite, by E2,
we have $\Kaxc \vdash_\phi \phi_s \rimp E_G \neg \phi_t$, as desired.

Returning to the proof of (\ref{eq5}), note that
(since $E_G \psi \in s$) if $\neg \psi \in t$ then $(s,t) \notin
\union_{i \in G} \K_i$.  Thus, from (\ref{eq2}) and
(\ref{basic3}),
we have
\begin{equation}\label{eq4.5}
\Kaxc \vdash_\phi \phi_s \rimp E_G(\land_{\{t \in S^j: \neg \psi \in t\}}
\neg \phi_t).
\end{equation}
By the induction hypothesis,
for all states in $t \in S^1 - S^j$, we have that $\Kaxc \vdash_\phi
\neg
\phi_t$.  Thus, using (\ref{eq-1}), we have
\begin{equation}\label{eq4}
\Kaxc \vdash_\phi \psi \dimp \land_{\{t \in S^j: \neg \psi \in t\}}
\neg\phi_t.
\end{equation}
(\ref{eq5}) now follows from (\ref{basic4}), (\ref{eq4.5}),
and (\ref{eq4}).

{F}inally, we must show that if $\neg C_G \psi \in s$ and
there is no state $t \in S^j$ $G$-reachable from $s$ in $S^j$ such that
$\neg \psi \in t$, then $\Kaxc \vdash_\phi\phi_s \rimp C_G \psi$, again
showing that $\phi_s$ is $\Kaxc$-inconsistent.  This follows by a
relatively straightforward modification of the completeness proof
given in \cite{FHMV,HM2}, so we just sketch
the details here.   Let $T_1 = \{ t \in S^j : \neg C_G \psi \in t$ and
there is no
state $t' \in S^j$ $G$-reachable from $t$ in $S^j$ such that $\neg \psi
\in t'\}$ and $T_2 =  \{t \in S^j: C_G\psi \in t\}$.  Let $T_i'$
consist of those states in $T_i$ that also contain $\psi$, $i = 1,2$.
Let $T = T_1 \union T_2$ and let $T' = T_1' \union T_2'$.
We claim that
there is no pair $(t,t') \in \union_{i \in G} \K_i$ such
that $t \in T$ and $t' \in S^j - T'$.  It is immediate that if $t
\in T_2$ then (since $\psi \land C_G \psi \in t/E_G \subseteq t'$)
$t' \in T_2'$.  If $t \in T_1$ and $t' \in S^j - T'$, then
either $\neg \psi \in t'$ or $\neg C_G \psi \in t'$ and there is a state
$t''$ $G$-reachable from $t'$ in $S^j$ such that $\neg \psi \in t''$.
This means that either $t'$ or $t''$ is a state $G$-reachable from
$t$
in $S^j$ containing $\neg \psi$.  This contradicts the fact that
$t \in
T_1$.

It now follows from
(\ref{eq2})
that for all $t \in T$ and
$t' \in S^j - T'$, we have
\begin{equation}\label{eq6.5}
\Kaxc \vdash_\phi\phi_{t} \rimp E_G \neg \phi_{t'}.
\end{equation}
Let $\phi_T = \lor_{t \in T} \phi_{t}$ and let $\phi_{T'} = \lor_{t'
\in T'} \phi_{t'}$.  By propositional reasoning, we have
$\Kaxc \vdash_\phi
\phi_{T'} \dimp (\phi_T \land \psi)$.
It easily follows from
(\ref{basic3}), (\ref{basic4}), and (\ref{eq6.5}) that $\Kaxc \vdash_\phi
\phi_{t} \rimp E_G
\phi_{T'}$.  Since this is true for all $t \in T$, we have
\begin{equation}\label{eq7}
\Kaxc \vdash_\phi\phi_T \rimp E_G(\phi_T \land \psi).
\end{equation}
By applying RC1 and the fact that $s \in T$, we have
$\Kaxc \vdash_\phi\phi_s \rimp C_G\psi$.  Since $\neg C_G \psi \in s$, it
follows that $\phi_s$ is $\Kaxc$-inconsistent.

This completes the completeness proof in the case of $\MP$.  To deal
with $\MPr$, we must just show that if $s$ is eliminated because $(s,s)
\notin \K_i$ for some $i \in\A$, then $\Taxc\vdash_\phi \neg \phi_s$;
all other cases are identical.  But if $(s,s) \notin \K_i$, then there
must be some $G$ and $\psi$ such that $i \in G$, $E_G \psi \in s$, and
$\neg \psi \in s$.  Since $\Taxcp$ includes the axiom $E_G \psi \rimp
\psi$, we have that
$\Taxcp \vdash_\phi \neg \phi_s$, as desired.
\eprf

\bigskip

\noindent
{\bf Proof of Theorem~\ref{complete} for $\MP$ and $\MPr$:}  We have
already observed that the axioms are sound.  For completeness,
suppose that $\phi$ is valid with respect to $\MP$.  By
Proposition~\ref{trans}, so is $\phi^{\sigmak}$.  By Lemma~\ref{proof},
there is a proof of $\phi^{\sigmak}$ in $\KaxcG$ that mentions only the
modal operators in $\phi^{\sigmak}$.
Given a formula $\psi$ in which the only modal operators that appear are
modal operators that appear in $\phi^{\sigmak}$ (and thus have the form
$E_{\sigmak(G)}$, $C_{\sigmak(G)}$, and $K_{\sigmak(i)}$, for sets $G$
and $\{i\}$ in $\G_\phi$) let $\psi^{\tauk}$ be the unique formula all
of whose modal operators appear in $\phi$ such that
$(\psi^{\tauk})^{\sigmak} = \psi$.  Lemma~\ref{replace} assures us that
$\psi^{\tauk}$ is well defined.
We can
pull the proof of $\phi^{\sigmak}$ back to a proof of $\phi$, by
replacing
each occurrence of a formula $\psi$ in the proof by $\psi^{\tauk}$.

The argument for $\MPr$ is identical, except that the proof
uses
instances of the axiom E5. These can be eliminated by using E1 and K2,
as we observed earlier (although now the proof of $\phi$ may use modal
operators $K_i$ that do not appear in $\phi$).  \eprf

\subsection{Dealing with $\M_\A^{\it rt}$}
Proposition~\ref{trans} as it stands does not hold for $\MPrt$.
There is no
guarantee that the translated formula is satisfiable in
$\MPrt$, even if $\phi$ is.  Indeed, suppose that $\G$ is closed under
intersection and complementation, so that we can use the function
$\sigma$ of Proposition~\ref{Shore}.  Suppose that $\phi$ is the
formula $E_G p \land \neg E_G E_G p$, where $|G| \ge 2$.  The formula
$\phi^{\sigma}$
looks syntactically identical, except that
$\sigma(G)$
is
a single agent in $\A'$.  We cannot make the $\K_G$ relation transitive
and still satisfy
$\phi^{\sigma}$.
More generally, to deal with
$\MPrt$,
we must be careful in how we deal with singleton sets.


As a first step, we define {\em mixed\/} structures.
Since we also
need these to deal with $\MPrst$ and $\MPelt$, we define three types of
mixed structures at once.  We say that a binary relation $\K$ is {\em
secondarily reflexive\/} \cite{Chellas} if $(s,t) \in \K$ implies $(t,t)
\in \K$.
Let $\MPrth$
%
(\respc $\MPrsth$; $\MPelth$)
consist of structures $M = (S,\pi, \{\K_i: i \in
\A_1 \union \A_2\})$  where the relations $\K_i$ for $i \in \A_1$
are reflexive and transitive
%
(\respc reflexive, symmetric and transitive; Euclidean, serial and transitive)
and the relation $\K_i$ for $i \in \A_2$ are reflexive
%
(\respc reflexive and symmetric; serial and secondarily reflexive).

We can now define our translation in the case of
$\MPrt$.
Although we can in fact get an analogue to Proposition~\ref{trans} for
$\MPrt$,
it turns out to be easier to provide a translation that combines
Proposition~\ref{trans} and Lemma~\ref{K}, rather than separating them.
As suggested by Proposition~\ref{reduction}, the translation involves
$\R(\G_\phi^1)$, rather than $\R(\G_\phi)$.
Given a formula $\phi$,
let $\A^{\phi,rt} = \{\H : \exists G [(G,\H) \in \R(\G_\phi^1)]\}$.  Let
$\A_1 = \{\H: \exists G [(G,\H) \in \R(G_\phi^1), \, |G- \union \H| =
1]\}$; let $\A_2 = \A^{\phi,rt} - \A_1$.
Define $\sigmart: \A \rightarrow \A^{\phi,rt}$
as before: $\sigmart(i) = \H$ if $i \in A_\H$
and $\sigmart(i)$
is undefined
otherwise.
Given $\H \in \A^{\phi,rt}$, we define
$\taurt(\H) = \inter (\G_\phi^1 - \H)$.  Since it is easy to see that
$\R(\G_\phi^1) = \R(\G_\psi)$ for some
appropriate
$\psi$, it
is immediate that Lemma~\ref{K} applies
to
$\sigmart$ and $\taurt$.

\pro\label{transrt}
$\phi$ is satisfiable in $\MPrt$ iff $\phi^{\sigmart}$
is satisfiable in $\MPrth$.
\epro

\prf
{F}irst suppose that $(M,s)\models\phi$, where $M \in \MPrt$. We convert
$M = (S,\pi, \{\K_i: i \in \A\})$
into a structure $M' = (S,\pi, \{\K_\H: \H \in \A^{\phi,rt}\})$ as
before, by defining $\K_{\H} = \union \{\K_i: i \in
\taurt(\H)\}$.  Since
Lemma~\ref{K} applies, the proof that
$(M',s) \sat \phi$ is identical to that in Proposition~\ref{trans}.  We
must only show that $M' \in \MPrth$.  Since the union of reflexive
relations is reflexive, it is immediate that $\K_{\H}$ is reflexive for
$\H \in \A_2$.  If $\H \in \A_1$, then
$|A_\H| = 1$.
Suppose that $A_\H = \{i\}$.  We claim that $\taurt(\H) = \{i\}$.
By construction, $\{i\} \in \G_\phi^1$.  We cannot
have $\{i\} \in \H$, since $i \notin \union \H$.  Thus
$\{i \} \in \G_\phi^1 - \H$, so
$\taurt(\H) =
\inter(\G_\phi^1 - \H) \subseteq \{i\}$.  Since $\taurt(\H) \ne \emptyset$
by Lemma~\ref{K}(c), we must have $\taurt(\H) = \{i\}$.  Thus, $\K_\H =
\K_i$, so $\K_\H$ is reflexive and transitive.

{F}or the opposite direction we need to work a little harder than before,
because we must ensure that all the $\K_i$ relations are reflexive and
transitive for all $i \in \A$.   Supppose
$(M,s)\models\phi^{\sigmart}$
for some $M = (S,\pi, \{\K_\H: \H \in \A^{\phi,rt}\}) \in \MPrth$.
Let $S_0$ and $S_1$ be two disjoint copies of $S$.  For a state $s \in
S$, let $s_i$ be the copy of $s$ in $S_i$, $i = 0,1$.
Let $M' = (S',\pi', \{\K_i: i \in \A\})$ be defined as follows:
\begin{itemize}
\item $S' = S_0 \union S_1$.
\item $\pi'(s_i) = \pi(s)$ for $i = 0,1$.
\item
If $\sigmart(i) \in \A_1$, define $\K_i = \{(s_i, t_j): (s,t) \in
\K_{\sigmart(i)}, i, j \in \{0,1\}\}$.   $\K_i$ is clearly reflexive and
transitive in this case, since $\K_{\sigmart(i)}$ is.
\item If $\sigmart(i) = \H \in \A_2$,
note that $|A_\H| \ge 2$.
It is immediate from the definition that $\sigmart(i) = \H$ for all
$i \in A_\H$.  Pick some $i_\H \in A_\H$.  If $i = i_\H$,
then define
$\K_{i}=\{(s_0,t_1): (s,t) \in \K_\H\} \union \{(s_j,s_j): j \in
\{0,1\}\}$; if $i \ne i_\H$, define $\K_i
= \{(s_1,t_0): (s,t) \in \K_\H\} \union \{(s_j,s_j): j \in
\{0,1\}\}$.  Clearly $\K_i$ is reflexive and transitive.
\end{itemize}
This construction guarantees that
\begin{equation}\label{eq8}
\mbox{$(s,t)
\in \K_\H$ iff
$(s_0,t_1), (s_1,t_0) \in \union_{\{i:
\sigmart(i) = \H\}}\K_i$}
\end{equation}
and
\begin{equation}\label{eq9}
\mbox{$(s_1,t_0) \in \union_{\{i: \sigmart(i) = \H\}}\K_i$ iff
$(s_0,t_1) \in \union_{\{i: \sigmart(i) = \H\}}\K_i$.}
\end{equation}

A straightforward argument by induction on structure now
shows that if $\psi \in {\cal L}_{\G_\phi^1}^C$, then the following are
equivalent for all $t \in S$:
\begin{itemize}
\item $(M,t) \sat \psi^{\sigmart}$,
\item both $(M',t_0) \sat \psi$ and $(M',t_1) \sat \psi$,
\item $(M',t_0) \sat \psi$ or $(M',t_1) \sat \psi$.
\end{itemize}
Of course, the interesting cases are if $\psi$ is of the form $K_i
\psi'$, $E_G \psi'$, or $C_G\psi'$.  These follow immediately from
observations (\ref{eq8}) and (\ref{eq9}).
\eprf

The next step is to get an analogue of Theorem~\ref{dec} for
$\MPrth$.
The basic idea of the proof is the same as that of
Theorem~\ref{dec}.  However, in our construction, we need to make
the $\K_i$ relations transitive.  To see the difficulty, suppose that
$\phi$
is $K_1 p \land E_G q$, where $G$ is a set of agents containing 1.
Recall that in Theorem~\ref{dec}, states are consistent subsets of
$\Sub^+(\phi)$.
Let $s$, $t$, and $u$ be states such that $s= \{K_1 p, E_G q, p, q\}$,
$t = \{K_1 p, \neg E_G q, p, q\}$, and $u = \{K_1 p, \neg E_G q, p, \neg
q\}$.  With our previous construction, we would have both $(s,t) \in
\K_1$
and $(t,u) \in \K_1$.  By transitivity, we should also have $(s,u)
\in \K_1$.  But since $E_G q \in s$ and $\neg q \in u$, we have
$(s,u) \notin \K_1$.   Nevertheless,
each of $s$, $t$, and $u$
individually seems consistent.
Which
state should we eliminate in order to preserve transitivity?

To deal with this problem, we need to put more information
(i.e., more formulas) into each state.  Intuitively, if $(s,t) \in
\K_i$, then we should have $K_i q \in t$, because if $E_G q \in s$, then
$K_i q$ should also be in $s$, as should $K_i K_i q$ by K4.  It would
then follow that $K_i q$ should be in $t$.  This, in turn, would
guarantee that $(t,u) \notin \K_i$, since $q \notin u$.

What we would like to do now is to augment $\Sub(\phi)$ by including all
formulas $K_i \psi$ such that $E_G \psi \in \Sub(\phi)$ and $i \in G
\inter \A_1$.  (We restrict to $\A_1$ since these are the only relations
that are required to be transitive.)
While this approach can be used to force the $\K_i$ relations
to be transitive, the resulting set of formulas can
have size
$O(|\A_1||\phi|)$, which means the resulting state space (the analogue
of $S^1$)
could then
have size $2^{|\A_1||\phi|}$.  This would not give us the desired
complexity bounds. Thus, we must proceed a little more cautiously.

\thm\label{decrt}
If $\A = \A_1 \union \A_2$ is finite and there is an algorithm for
deciding if $i
\in G$ for $G \in \G$ that runs in time linear in $|\A|$, then
there is a constant $c > 0$ (independent of $\A$) and
an algorithm
that, given a formula
$\phi$ of $\LGC$, decides if $\phi$ is satisfiable
in
$\MPrth$
and runs in time $O(|\A|2^{c|\phi|})$.
\ethm
\prf
We assume for ease of exposition that $\A_1 \ne \emptyset$; we leave the
straightforward modification in case $\A_1 = \emptyset$ to the reader.
{F}or each $i \in \A_1$, let $\ESub_i(\phi)$
be the least set containing $\Sub(\phi)$ such that if $E_G \phi \in
\Sub(\phi)$ and $i \in G$, then $K_i \phi \in
\ESub_i(\phi)$.
It is easy to see that
$|\ESub_i(\phi)| \le 2|\Sub(\phi)|$, since we add at most
one formula for each formula in $|\Sub(\phi)|$.
Let $S^1_i$ consist of all the subsets of $\ESub^+_i(\phi)$ that are
maximally consistent, and now let $S^1 = \union_{i \in \A_1} S^1_i$.
Note that, as modified, $|S^1| \le 2^{2|\phi|}$.  Thus,
this modification
keeps us safely within the
desired exponential time bounds.

We keep the definition of $\K_i$ unchanged for $i \in \A_2$
(\ie $(s,t) \in \K_i$ iff $s/\overline{K}_i \subseteq t$),
but we
need to modify it for $i
\in \A_1$.  We redefine $\K_i$ for $i \in \A_1$ by defining
$(s,t) \in \K_i$ iff
$s/\oKi \union \{K_i \psi: K_i \psi \in s\} \subseteq
t \inter (t/\oKi \union \{K_i \psi: K_i \psi \in t\})$.
It is easy to check that this modification forces the $\K_i$
relations to be transitive.  We force all the $\K_i$ relations to be
reflexive just as with $\MPr$, by eliminating $s \in S^1$ if
$(s,s) \notin \K_i$ for some $i \in \A_1 \union \A_2$.
The remainder of the construction---eliminating the states that do not
seem consistent---is unchanged.




We now need to show that the algorithm is correct.  First suppose that
$\phi$ is satisfiable in $\MPrth$.
In that case, $(M,s_0)\sat \phi$ for
some structure $M = (S,\pi, \{\K_i': i \in \A_1 \union \A_2\})
\in \MPrth$.  We can associate with each state $s \in S$ and $i \in
\A_1$ the state $s^*_i$ in
$S^1_i$ consisting of all the formulas $\psi \in
\ESub_i(\phi)$
such that
$(M,s) \sat \psi$.  It is easy to see that if $(s,t) \in \K_i'$ then
$(s^*_j,t^*_i) \in \K_i$ for all $j$.%
\footnote{Note that it is not necessarily the case
that $(s^*_j,t^*_{j'}) \in \K_i$ for $j' \ne i$.  For example, suppose
$\phi$ is the formula $E_G p$, $i \in G \inter \A_1$, and $M$ is such
that $(M,s)
\sat E_G p \land p$, $(M,t) \sat \neg E_G p \land p$, and $(s,t) \in
\K_i$.  Then for $i \ne j, j'$ and
$j\notin G$, we have $s^*_j = \{E_G
p, p\}$ and $t^*_{j'} = \{p,\neg E_G p\}$.  Since $p \in s^*_j/\oKi -
t^*_{j'}/\oKi$, we have that $(s^*_j,t^*_{j'}) \notin \K_i$.}
Using this observation, a straightforward
induction shows that the states $s^*_i$ for $s \in S$ always seem
consistent, and thus are in $S^j$ for all $j$ and all $i \in
\A_1$.  Moreover, $\phi \in
(s_0)^*_i$ for all $i \in \A_1$.  Thus, the algorithm will declare
that $\phi$ is satisfiable, as desired.

Conversely, suppose that the algorithm declares that $\phi$ is
satisfiable.
We construct a structure $M = (S,\pi, \{\K_i': i \in \A_1 \union \A_2\})
\in \MPrth$ in which $\phi$ is satisfied just as
Theorem~\ref{dec}.  Our modified construction
guarantees that the $\K_i'$ relations are all reflexive and the ones in
$\A_1$ are transitive.  \eprf


We are almost ready to prove Theorem~\ref{complexity} for $\MPrt$.
However, we first we need to characterize the complexity of
translating from $\phi$ to $\phi^{\sigmart}$.  In particular, we need a
bound on the number of elements in $\R(\G_\phi^1)$ and the
number of oracle calls required to compute them.
To do this, we first define two
auxiliary sequences of sets $\D_i^m(\J)$
and $\E_i^m(\J)$, $i = 1, 2, 3, \ldots$.  (We omit the parenthetical $\J$
when it is clear
from context.)   Fix $m$. Let $\D_0^m = \J$ and $\D_{i+1}^m =
\J \union
\{G - \union\H:
(G,\H) \in \R(\D_i^m) \mbox{ and } |G - \union\H|\leq m\}$;
let $\E_i^m = \D_i^m - \D_0^m$.
Set $\D^m = \union_i \D_i^m$ and $\E^m = \union_i \E_i^m$.
{F}inally, denote $\R(\D^m)$ by $\R^m(\J)$.
It is easy to check that $\D_0^m \subseteq \D_1^m \subseteq \ldots$
and that $\R^m(\J) = \union_i \R(\D^m_i)$.

The next lemma provides partial motivation for these definitions.
\lem\label{equiv}
$\R^m(\J) = \R(\J^m)$.
\elem

\prf An easy induction on $i$ shows that $\D^m_i(\J) \subseteq \J^m_i$
(as defined in Definition~\ref{Gm}) for all $i$, so $\D^m\subseteq
\J^m$. We next show that
every set in $\J^m_i$ is the union of sets in $\D^m$, by induction on
$i$.  This is immediate if $i=0$, since $\J^m_0 = \D^m_0 = \J$.  Suppose
that the result holds for $\J^m_i$; we show it for $\J^m_{i+1}$.
Suppose that
$H \in \J^m_{i+1}$.  If $H \in \J$, then clearly $H \in \D^m$.  Thus,
without loss of generality, $H \in \J^m_{i+1} - \J$, which means that
$|H| \le m$.  Let $H'$ be the union of all sets in $\D^m$ contained in
$H$.  If $H' = H$, then we are done.  Suppose by way of contradiction
that $H
- H' \ne \emptyset$.  We obtain a contradiction to the choice of
$H'$ by showing that $H - H'$ contains a set in $\D^m$.

Since $H'$ is finite, it can be written as a finite union 
of 
sets in
$\D^m$, say of
$\H_1 = H_1, \ldots, H_k$.  Since $H \in \J^m_{i+1} - \J$,
$H = G - \union \H_2$ for
some $G \in \J$ and $\H_2 \subseteq \J^m_i$.  By the induction
hypothesis,
there exists some $\H_3 \subseteq \D^m$ such that $\union \H_2 = \union
\H_3$.  There must exist some set $\H_4 \supseteq \H_1
\union \H_3$ such that $(G, \H_4) \in \R(\D^m)$.  But then $H - H'
\supseteq G - \union \H_4 \in \D^m$, and we obtain the desired
contradiction.

It now easily follows that $\R(\J^m) = \R(\D^m) = \R^m(\J)$. \eprf

The following result will be used to help compute the elements of
$\R^m(\J)$.

\lem\label{Gmcount}
Let $\J$ be a set of subsets of $\A$ with $|\J| = n$.
\begin{itemize}
\item[(a)] If $(G,\H) \in \R(\D)$, where $\J \subseteq \D \subseteq
\J^*$, then $G - \union \H$ is an atom over $\J$.
\item[(b)] $\J \subseteq \D_i^m \subseteq \J^*$ for all $i, m$.
\item[(c)] $|\{\H: \exists G \in \D^m((G,\H) \in \R^m(\J)\}| \le 2^n$.
\item[(d)]  If $(G,\H) \in \R(\D_i^m)$, then
either $G \in \J$ and $\E_i^m\subseteq \H$ or
$A_\H \in \E_i^m$ and $\E_i^m - \{A_\H\} \subseteq \H$.
Moreover,
if $(G,\H) \in \R^m(\J)$, then
either $G \in \J$ and $\E^m\subseteq \H$ or $(G - \union\H) \in
\E^m$ and $\E^m - \{G- \union \H\} \subseteq \H$.
\item[(e)]  $\D^m = \D_n^m$ and $\E^m = \E_n^m$.
\end{itemize}
\elem

\prf
{F}or part (a), we know from Lemma~\ref{unique}(a) that if $(G,\H) \in
\R(\D)$, then $G - \union \H$ is an atom over $\D$.  Since
$\J \subseteq \D \subseteq
\J^*$, it is immediate that it must in fact be an atom over $\J$ as
well.
(Recall that $\J^*$ is the algebra  generated by $\J$.)

Part (b) follows immediately from (a),
since an easy induction on $i$ shows that $\E_i^m \subseteq \J^*$.

For part (c), by Lemma~\ref{unique}(a), it follows that
$A_\H$ is an atom over
$\D^m$.  But since $\J \subseteq \D^m = \union_i \D_i^m \subseteq \J^*$
by part
(b), it follows that $A_\H$ is actually at atom over $\J$.  Moreover if
$(G',\H') \in \R^m(\J)$ and $\H \ne \H'$, then it follows from
Lemma~\ref{unique}(c) that $A_\H \ne A_{\H'}$.  Since there are at most
$2^n$ atoms over $\J$, part (c) follows.

{F}or part (d), if $(G, \H)
\in \R (D_i^{m})$ then,
by Lemma~\ref{unique}(a), $A_\H = G - \union \H$ is an atom over
$\D_i^m$ and has the form
$\cap (\D_i^m - \H)\cap\cap \{\overline H :H \in \H\} $.
By the arguments of part (c), $A_\H$ is also an atom over $\J$.
We say that the
sets in $\D_i^m - \H$ {\em appear positively\/} in $A_\H$ and the sets
in
$\H$ {\em appear negatively\/} in $A_\H$.
If one of the sets $G' \in \E_i^m$ appears positively in $A_\H$
then clearly $A_\H \subseteq G'$.
But since the elements of $\E_i^m$ are also atoms over $\J$, it follows
that in this case $A_\H = G' \in \E_i^m$ and,
since $\H$ is $G$-maximal,
$\E_i^m - \{A_\H\} \subseteq \H$.
Otherwise, $\E_i^m\subseteq \H$ as required;
moreover, since $\D_i^m = \E_i^m \union \J$ and $G \notin \H$, we must
have $G \in \J$.  The argument for the second half of (d) is identical.

Clearly the two claims in part (e) are equivalent. We prove the second.
As observed in the proof of (c), every set in
$\E^m$
is an atom $A$ over $\J$.
It is easy to see that there are no
atoms in
$\E^m$
where all $n$ sets in $\J$ appear negatively, since
every set in
$\E^m$
is a nonempty subset of some $G \in \J$.
(This can be proved by induction on $i$ for each $\E^m_i$.)
We prove by induction on $i$ that if
$A \in \E^m$ and $n-i$ sets appear negatively in $A$ for $i
\ge 1$, then $A \in \E^m_{i}$.

Clearly if $i=1$, then $A = G - (H_1 \union \ldots \union H_{n-1})$,
and $\H = \{H_1, \ldots, H_{n-1}\}$ is a $G$-maximal subset of $\J$.
Thus, $(G,\H) \in \D_1^m$ and $A \in \E_1^m$.  Suppose that the
result is true if $i=k$ and suppose that $n-(k+1)$ sets appear
negatively in $A$. As $A \in \E^m$,
there must be some minimal $j$ such that
$A \in \E_{j+1}^m$.
By definition, $A = \A_H$ for some $(G,\H) \in \R(\D_j^m)$.
By (d), either
$A = G - (\union\H' \union \E_j^m)$ and $\H' \subseteq \J$ or $A \in
\E_j^m$.
The latter case contradicts our choice of $m$,
so we may assume that
$A = G - (\union\H' \union
\E_j^m)$ and $\H' \subseteq \J$.
It is easy to see that $\H'$ must
consist of precisely the sets in $\J$ that appear negatively in $A$.
(If it did not include all the sets that appear negatively in $A$
then $\H'
\union \E_j^m$ would not be a $G$-maximal subset of $\J \union \E_j^m$;
if it includes any sets that appear postively then $A$ would be
empty.)  Let $\E'$ consist of all the atoms $A'$ in $\E_j^m$ in
which the set of sets in $\J$ that appear negatively in $A'$ is a
strict superset of
$\H'$.  It is easy to see that $G - (\union\H' \union \E_j^m) =
G - (\union\H'
\union \E')$, since all the sets in $\E_j^m - \E'$ must be disjoint
from $G - \union\H'$.
(This is clear for the $B \in \E_j^m - \E'$ for which some set appearing
negatively in
$A$ does not appear negatively in $B$.  On the other hand, if the same
sets appear negatively in $B$ as in $A$ then $B=A$ and we contradict
the minimality of $j$.)
By the induction hypothesis, $\E' \subseteq
\E^m_{n-k}$.  Thus,
$A = G - (\H' \union \E_{n-k}^m) \in \E_{n-k+1}^m$,
as desired.  \eprf

We remark that a simpler proof, just using the fact that there are at
most $2^n$ atoms over $\J$, can be used to show that $\E^m_{n'} =
\E^m_{2^n}$ for $n' > 2^n$.  This simpler proof would suffice for the
purposes of this subsection.  However, we shall use the added
information in (e) in Section~\ref{oracle}.

\bigskip

\noindent
{\bf Proof of Theorem~\ref{complexity} for $\MPrt$:}
Again, the lower bound follows from standard results in \cite{HM2}.

{F}or the upper bound, suppose that
we are given a formula $\phi$ such that $n
= |\phi|$
%
and $\H \in \A^{\phi,rt}$.  By definition, there exists a
$G$ such that $(G,\H) \in \R(\G_\phi^1)$.
By Lemma~\ref{equiv}, $\R^1(\G_\phi) = \R(\G_\phi^1)$.
Thus, $\H \subseteq \D^1(\G_\phi) = \G_\phi \union E^1_n(\G_\phi)$.
By Lemma~\ref{Gmcount}(d), either $\E^1_n(\G_\phi) \subseteq \H$ or
$\H$ contains all but one element of $\E^1_n(\G_\phi)$.  Thus,
we can uniquely characterize $\H$ by a pair
$(\H',X)$, where
$\H' = \H \inter \G_\phi$
and $X = \E^1_n(\G_\phi) - \H$ (so that $X$ is either the empty set or a
singleton).
It should be clear that we can compute
compute the set $\E^1_{n}(\G_\phi)$ in time $O(n^2 2^{cn})$ and
which of these (at most $2^{2n} + 2^n$)
pairs
is in $\A_1$ and
$\A_2$ using at most $2n(2^{2n} + 2^n)$ calls to the oracle $O_1$.

By
Lemmas~\ref{K}(a) and~\ref{Gmcount}, we can
similarly
compute the
formula $\phi^{\sigmart}$ in time
$O(2^{cn})$ using $O(2^{cn})$ oracle calls.  We now
apply Proposition~\ref{transrt} and Theorem~\ref{decrt}, just as we
applied Proposition~\ref{trans} and Theorem~\ref{dec} in the case of
$\MP$.  \eprf

We next want to prove Theorem~\ref{complete} for $\MPrt$.  Just as with
$\MP$ and $\MPr$, we want to pull a proof of $\phi^{\sigmart}$ back to
a proof of $\sigma$.  However, it is no longer true that we can
necessarily prove $\phi^{\sigmart}$ using only the modal operators that
appear in $\phi^{\sigmart}$.  We may also need to use $K_\H$ for $\H \in
\A_1$.  Fortunately, this does not cause us problems.  The following
extension of Lemma~\ref{replace} is immediate.

\lem\label{replacert} The mapping $\sigmart$ (when viewed as a map with
domain $2^\A$) is injective on
$\G_\phi^1$.
\elem

Let $\fouraxcp$ consist of the axioms in $\Taxcp$ (so that, in
particular, E5 is included), together with every instance of K4 ($K_i
\phi \rimp K_i K_i \phi$) for $i \in \A_1$.  We write $\fouraxcp
\vdash_\phi \psi$ if there is a proof of $\psi$ in $\fouraxcp$ using
only the modal operators that appear in $\phi$ and $K_i$ for $i \in
\A_1$.

\lem\label{proofrt}  If $\A$ is finite and $\phi \in \LGC$ is
valid with respect to $\MPrth$, then $\fouraxcp
\vdash_\phi \phi$.   \elem

\prf The proof is similar to that of Lemma~\ref{proof} for
$\MPr$, except
that since the definition of the $\K_i$ relation is different, we must
still check that the results still hold with the modified definition.


Suppose that $s \in S^j$ does not seem consistent because $\neg E_G \psi
\in s$ and there is no state $t \in S^j$ such that
$(s,t) \in \union_{i \in G}\K_i$ and $\neg \psi \in t$.
We want to show that $\fouraxcp \vdash_\phi \phi_s \rimp E_G \psi$.
As before this suffices.

{F}or each $i \in G$ and, by induction, each $j$, we have a provable
equivalence for $\psi$ similar to the one
before: $\fouraxcp \vdash_\phi \psi
\dimp \land_{\{t \in {S_i}^j): \neg\psi \in t\}} \neg\phi_t$.  So it suffices
to find, for each
such $i$ and each $t \in S^j_i$ with $\neg\psi \in t$,
a $G^{i,t}$ containing $i$ such that $\fouraxcp \vdash_\phi
\phi_s \rimp E_{G^{i,t}} \neg\phi_t$.
{F}or $i \in \A_2$, this follows just as before.
{F}or $i \in \A_1$, we show that $\fouraxcp \vdash_\phi
\phi_s \rimp K_i \neg \phi_t$.
By our assumption $(s,t) \notin \K_i$.
Thus, there exists some formula $\theta \in s/\oKi \union \{K_i
\theta: K_i \theta \in s\} - (t \inter (t/\oKi \union \{K_i \theta: K_i \theta
\in t\}))$. If $\theta \in s/\oKi$, then
$\fouraxcp \vdash_\phi \phi_s \rimp K_i \theta$.
If $\theta = K_i \theta'$ is in $s$, then $\fouraxcp \vdash_\phi \phi_s \rimp
K_i \theta'$.  By K4, we have that $\fouraxcp \vdash_\phi \phi_s \rimp K_i
K_i \theta'$.  Thus, in either case, we have $\fouraxcp \vdash_\phi \phi_s
\rimp K_i \theta$.
  Since $\theta \in s/\oKi \union \{K_i \theta: K_i \theta \in s\}$,
it follows that $K_i \theta \in \ESub_i(\neg \phi)$.  We cannot have $K_i
\theta \in t$, for then (since $(t,t) \in \K_i$, so $t/\oKi \subseteq t$)
we would have $\theta \in t \inter t/\oKi$, contradicting our  choice of
$\theta$.  Thus
we must have that $\neg K_i \theta \in t$.  It follows that $\fouraxcp
\vdash_\phi K_i \theta \rimp \neg \phi_t$.  Using (\ref{basic4}), we get
that $\fouraxcp \vdash_\phi K_i K_i \theta \rimp K_i \neg \phi_t$.
Since $\fouraxcp \vdash_\phi K_i \theta \rimp K_i K_i \theta$ and, as
shown earlier,
$\fouraxcp \vdash_\phi \phi_s \rimp K_i \theta$, it follows that
$\fouraxcp \vdash_\phi \phi_s \rimp K_i \neg \phi_t$, as desired.

{F}inally, we must show that if $\neg C_G \psi \in s$ and
there is no state $t \in S^j$ $G$-reachable from $s$ in $S^j$ such that
$\neg \psi \in t$, then $\fouraxc \vdash_\phi\phi_s \rimp C_G \psi$.
This argument is identical to that given in the proof of
Lemma~\ref{proof}, so we do not repeat it here.
\eprf

\bigskip

\noindent
{\bf Proof of Theorem~\ref{complete} for $\MPrt$:}  Again, we have
already observed that the axioms are sound.  For completeness, suppose
that $\phi$ is valid with respect to $\MP$.  By
Proposition~\ref{transrt}, $\phi^{\sigmart}$ is valid with respect to
$\MPrth$.
By Lemma~\ref{proofrt},
there is a proof of $\phi^{\sigmart}$ in $\KaxcG$ that mentions only the
modal operators in $\phi^{\sigmart}$ and the operators $K_\H$ for $\H
\in\A_1$.   Using Lemma~\ref{replacert}, it follows that we can
pull this back to a proof of $\phi$ in $\fouraxc$.  \eprf

\subsection{Dealing with $\M_\A^{\it rst}$}
$\MPrst$ and $\MPelt$ introduce additional complications.  The
translation used in Proposition~\ref{transrt} no longer suffices.
We need to deal with the fact that in $\MPrst$, we can test not only
that whether a set is a singleton, but whether it has size $k$ for any
$k$.
Given a formula $\phi$, suppose that $|\phi| = n$.
We want to map $\A$ to a finite set of agents and prove an analogue
of
Propositions~\ref{transrt}.
The obvious analogue of $\A^{\phi,rt}$ would be to consider the sets
$\H$ such that $(G,\H) \in
\R(\G_\phi^n)$.
We essentially do this,
except that we replace all sets of cardinality $\le n$ by the singletons
in them.

Given a set $\J$ of subsets of $\A$, let $\tJ^m = \D^m(\J)
\union \{\{i\}: \exists G \in \D^m(\J)
(|G| \le m, \, i \in G\}$.
Let $\A^{\phi,rst} = \{\H : \exists G ((G,\H) \in
\R(\tG_\phi^{n})\}$. Let
$\A_1 = \{\H: \exists G [(G,\H) \in \R(\tG_\phi^{n}), \, |G- \union
\H| = 1\}$; let $\A_2 = \A^{\phi,rst} - \A_1$.
Define $\sigmarst: \A \rightarrow \A_1 \union \A_2
$ as before: $\sigmarst(i) = \H$ if
$i \in A_\H$ and $\sigmarst(i)$
is undefined
otherwise.
Much
as before, we define
$\taurst(\H) = \inter (\tG_\phi^{n} - \H)$.  Since it is easy to
see that
$\R(\tG_\phi^{n}) = \R(\G_\psi)$ for some appropriately chosen
$\psi$, it
is immediate that Lemma~\ref{K} applies without change to
$\sigmarst$ and $\taurst$.

\lem\label{separation}
If $\H \in \A_2$, then $|A_\H| \ge n+1$. \elem
\prf Suppose, by way of contradiction, that $\H \in \A_2$ and $1 \le
|A_\H| \le n$.  We must have $|A_\H| > 1$, for otherwise $\H \in \A_1$.
Since $\A_2 \subseteq \A^{\phi,rst}$,
there must exist $G \in \D^n(\G_\phi)$ such that $\H$ is $G$-maximal.
But if $|A_\H| \le n$, then every singleton subset of $A_\H$ is in
$\tG_\phi^n$.  This contradicts the fact that $\H$ is $G$-maximal,
because if $\H'$ is $\H$ together with one of 
these singleton subsets, 
we must have $G - \union \H' \ne \emptyset$.  \eprf

\pro\label{transrst}
$\phi$ is satisfiable in
$\MPrst$
iff
$\phi^{\sigmarst}$
is satisfiable in $\MPrsth$.
\epro

\prf
{F}irst suppose that $(M,s)\models\phi$, where $M \in \MPrst$. We
convert
$M = (S,\pi, \{\K_i: i \in \A\})$
into a structure $M' = (S,\pi, \{\K_\H: \H \in \A^{\phi,rst}\})$ as
before, by defining $\K_{\H} = \union \{\K_i: i \in \taurst(A)\}$.
As the union of symmetric relations is symmetric,
the proof that this works is
essentially
identical to that in Lemma~\ref{transrt}
for the case of $\MPrt$.

{F}or the opposite direction,
suppose that
$(M,s)\models\phi^{\sigmarst}$
for some $M = (S,\pi, \{\K_\H: \H \in \A^{\phi,rst}\}) \in \MPrsth$.
We must
construct
a structure $M' \in \MPrst$ that satisfies $\phi$.
The state space for
the
structure $M'$ will again consist of copies of $S$,
but two copies no longer suffice to guarantee that the $\K_i$ relations
are equivalence relations.  In fact, we
use
countably many copies.

By Lemma~\ref{separation}, for each $\H \in \A_2$, there exist at least
$n+1$ agents in
$A_\H$.  Choose $n+1$ such agents, and call them $i_\H^0, \ldots,
i_\H^n$.
Partition $A_\H$ into $n+1$ disjoint sets $G_{\H,j}$
with $i_\H^j \in G_{\H,j}$.
We build copies of $M$ in a tree-like manner.
We index the copies of $M$ with strings of the form $((s_1,t_1), i_1,
\ldots, (s_k,t_k), i_k)$, such that $s_j,t_j \in S$, $i_j$ is
$i_\H^{j'}$ for some
$\H \in \A_2$ and $0 \le j' \le n$, $(s_j,t_j) \in \K_\H$,
and $i_j \ne i_{j+1}$.  Roughly speaking,
between $M_{\sigma}$ and $M_{\sigma\cdot((s_k,t_k),i_k)}$ we have
edges for the $\K_{i}$ relations for $\{i\} \in \A_1$ and also edges
between $s_{k}$ and $t_k$ in $\K_{i_k}$; however, there are no
edges in $\K_j$ if $\{j\} \notin \A_1$ and $j \ne i_k$; moreover, there
are no other edges in $\K_{i_k}$
except
those required
to assure reflexivity.

Before we can construct $M'$, we need some preliminary observations.  We
can suppose
that the states in $S$ are numbered.  Thus, for each state $s \in S$, if
$(M,s) \sat \neg C_G \psi$, there is a lexicographically minimal
shortest path $(s_0,
\ldots, s_k)$ such
that $(s_i, s_{i+1}) \in \K_\H$ for some $\H \in G$
 and
$(M,s_k) \sat \neg
\psi$.  Note that,
for each $i \leq k$, $(M,s_i) \sat \neg C_G \psi$ and
$(s_i, \ldots, s_k)$
is also the lexicographically
minimal shortest $G$-path from
$s_i$
leading to a state that satisfies
$\neg \psi$.  For each $s \in S$
and $B = E$ or $C$,
let
$\neg B_{G_1} \psi_1,\ldots, \neg B_{G_k} \psi_k$
be the formulas in $Sub^+(\phi)$ such that
$(M,s) \sat (\neg B_{G_j} \psi_j)^{\sigmarst}$.
{F}or each state $s \in
S$, we can associate a set $F(s)$ of at most $n$ pairs $(\H,t)$ such
that $(s,t) \in \K_\H$ and for every formula
$B_{G}
\psi \in \Sub(\phi)$, if $(M,s) \sat
(\neg B_G
\psi)^{\sigmarst}$, then there exists a pair $(\H,t) \in F(s)$ such that
$t$ is the first state
after $s$ on the lexicographically minimal $\sigmarst(G_j)$-path from
$s$
to a state satisfying $\neg\psi$.

We can now define a set
$\Sigma$ of strings inductively.  Let $\Sigma_0$ be the empty string.
Suppose that we have constructed $\Sigma_k$ consisting of strings
$((s_1,t_1), i_1, \ldots, (s_k,t_k), i_k)$ with the properties given
above.
{F}or each $\sigma = ((s_1,t_1), i_1, \ldots, (s_k,t_k), i_k)
\in \Sigma_k$, $s \in S$,
$(\H,t)
\in F(s)$, such that $\H \in
\A_2$, there is
exactly one string $\sigma \cdot ((s,t),i) \in \Sigma_{k+1}$.  We choose
$i \in A_\H$ in such a way that $i \ne i_k$, $i$ is one of $i_\H^0,
\ldots, i_\H^n$,  and a different $i$ is chosen for each
$(\H,t)
\in
{F}(s)$.  Since $|F(s)| \le n$ and we can choose among $n+1$ agents
$i_0^\H, \ldots, i_n^\H$, this can clearly be done. Let $\Sigma =
\union_k \Sigma_k$.

Let $M' = (S',\pi', \{\K_i: i \in \A\})$ be defined as follows:
\begin{itemize}
\item $S' = \union_{\sigma \in \Sigma} S_\sigma$, where
each $S_\sigma$ is 
a disjoint copy of $S$. 
We denote by $s_\sigma$ the copy of state
$s \in S$ in
$S_\sigma$.
\item $\pi'(s_\sigma) = \pi(s)$ for $s \in S$, $\sigma \in
\Sigma$.
\item
If $\sigmarst(i) \in \A_1$, define $\K_i = \{(s_\sigma, t_{\sigma'}):
(s,t) \in\K_{\sigmarst(i)}, \sigma, \sigma' \in \Sigma\}$.   $\K_i$ is clearly
reflexive, symmetric, and transitive in this case, since
$\K_{\sigmarst(i)}$ is.
\item If
$\sigmarst(i) =
\H \in \A_2$
and $i \in G_{\H,j}$,
then $\K_i =
\{(s_\sigma,s_\sigma): s \in S, \sigma \in \Sigma)\} \union
\{(s_\sigma,t_{\sigma'}),(t_{\sigma'},s_\sigma): \sigma' = \sigma\cdot
((s,t),i^j_\H) \mbox{ and } (s,t) \in\K_{\sigmarst(i)}\}$.
Again, it is clear from the construction that $\K_i$
is reflexive, symmetric, and transitive.
\item If $\sigmarst(i)$ is undefined, then $\K_i =
\{(s_\sigma,s_\sigma): s \in S, \sigma \in \Sigma)\}.$
Of course, in this case $\K_i$
is also reflexive, symmetric, and transitive.
\end{itemize}

We claim that for each formula $\psi \in Sub^+(\phi)$, the following are
equivalent:
\begin{itemize}
\item[(a)] $(M,s) \sat
\psi^{\sigmarst}$,
\item[(b)] $(M',s_\sigma) \sat \psi$ for all $\sigma \in \Sigma$,
\item[(c)] $(M',s_\sigma) \sat \psi$ for some $\sigma \in \Sigma$.
\end{itemize}
The argument proceeds by a straightforward induction on the structure of
$\psi$.  The argument that (a) implies (b) is easy using the induction
hypothesis, and the implication from (b) to (c) is trivial.  For the
argument that (c) implies (a), the only interesting cases are when
$\psi$ is of
the form $K_i \psi'$, $E_G \psi'$ or $C_G \psi'$.  For $K_i \psi'$, the
argument is easy because it is easy to see that $\{i\} \in \A_1$.  For
$E_G \psi'$,
suppose that $(M',s_\sigma)
\sat E_G \psi'$.  Then we must have $(M,s) \sat (E_G
\psi')^{\sigmarst}$.  For suppose not.  Then
there is some
$(\H,t)
\in F(s)$ such that $\H \in \sigmarst(G)$ and
$(s,t) \in \K_\H$.  Our construction guarantees that
$\sigma' = \sigma\cdot ((s,t),i)
\in \Sigma$ for some $i \in A_\H$.  From Lemmas~\ref{unique}(a)
and~\ref{K}(a), it follows that $i \in G$.  Moreover, by our
construction, $(s_\sigma,t_{\sigma'}) \in \K_i$.  The induction
hypothesis
now
guarantees that $(M',t_{\sigma'}) \sat \neg \psi'$.  But this
contradicts the assumption that $(M',s_\sigma) \sat E_G \psi'$.

{F}inally, suppose that $(M',s_\sigma) \sat C_G \psi'$.  Again, for a
contradiction, suppose that $(M,s) \sat \neg (C_G \psi')^{\sigmarst}$.
Now we proceed by a subinduction on the length of the shortest
$\sigmarst(G)$-path
in $M$ leading to a state satisfying $(\neg \psi')^{\sigmarst}$ to show
that $(M',s_\sigma) \sat \neg C_G \psi'$.   We leave
the straightforward details to the reader.
\eprf

Next, we want an analogue of Theorem~\ref{decrt} for $\MPrst$.  The
reader will not be surprised to learn that there are new complications
here as well, although the basic result still holds.

\thm\label{decrst}
If $\A = \A_1 \union \A_2$ is finite and there is an algorithm for
deciding if $i
\in G$ for $G \in \G$ that runs in time linear in $|\A|$, then
there is a constant $c > 0$ and
an algorithm that, given a formula
$\phi$ of $\LGC$, decides if $\phi$ is satisfiable
in
$\MPrsth$
and runs in time $O(|\A|2^{c|\phi|})$
\ethm

\prf  We start as in the proof of Theorem~\ref{decrt}.  Again, we assume
for ease of exposition that $\A_1 \ne \emptyset$.  For $i \in \A_1$,
let $S^1_i$ consist of all the subsets of $\ESub^+_i(\phi)$
that are maximally consistent and let $S^1 = \union_{i \in \A_1} S^1_i$.
The definition of the $\K_i$ relations depends on whether $i \in \A_1$
or $i \in \A_2$.  For $i \in \A_1$, we define the $\K_i$ relations on
$S^1$ so that $(s,t) \in \K_i$ iff
$s/\oKi \union \{K_i \psi: K_i \psi \in s\} \subseteq t$ and
$s/\oKi \union \{K_i \psi: K_i \psi \in s\} =
t/\oKi \union \{K_i \psi: K_i \psi \in t\}$.
It is easy to check that this modification forces these $\K_i$
relations to be Euclidean and transitive.
{F}or $i \in \A_2$ we define $\K_i$ so that $(s,t) \in \K_i$ iff
$s/\oKi \subseteq t$ and $t/\oKi \subseteq s$.
Clearly this modification forces these $\K_i$ relations to be symmetric.
We force all the $\K_i$ relations to be
reflexive just as with $\MPr$, by eliminating $s \in S^1$ if
$(s,s) \notin \K_i$ for some $i \in \A_1 \union \A_2$.


We now must also change the definition of $s$ seeming consistent.
Define the 
relations 
$\preceq_i$ on $S^1 \times S^1_i$ by taking
$s\preceq_{i}s^{\prime}$ if $s^{\prime}\in S_{i}^{1}$ and $s\cap
ESub_{i}(\phi)\subseteq s^{\prime}$. Suppose that we have defined $%
S^{1},\ldots,S^{m}$. $S^{m+1}$ consists of all states $s\in S^{m}$ that {\em %
seem consistent}, in that
the following three conditions hold (where we assume
that all states considered are in $S^{m}$):
\begin{enumerate}
\item  For all $i\in {\cal A}_{1}$, there exists an $s^{\prime }\in S^{j}$
such that $s\preceq _{i}s^{\prime }$.

\item  There exist distinct agents $i_{1},\ldots ,i_{k}\in {\cal A}_{1}$ and
states $s_{1},\ldots ,s_{k}$ such that $s\preceq _{i_{h}}s_{h}$ for $h\in
\{1,\ldots ,k\}$ and for every formula of the form $\lnot E_{G}\psi \in s$,
there is a $t$ such that either
\begin{enumerate}
\item  $(\exists i\in G\cap {\cal A}_{2})((s,t)\in {\cal K}_{i}\,\wedge
\,\lnot \psi \in t)$ or

\item  $(\exists h\leq k)(i_{h}\in G\,\wedge \,(s_{h},t)\in {\cal K}%
_{i_{h}}\,\wedge \,\lnot \psi \in t)$.
\end{enumerate}

\item  If $\lnot C_{G}\psi \in s$ then there exist states $%
s_{0},s_{0}^{\prime },s_{1},s_{1}^{\prime },\ldots ,s_{k}$ such that $s=s_{0}
$, $\lnot \psi \in s_{k}$, and there exist $j_{0},\ldots ,j_{k-1}$ in $G$
such that, for each $i\leq k$, $(s_{i}^{\prime },s_{i+1})\in {\cal K}_{j_{i}}
$ and either $j_{i}\in {\cal A}_{2}$ and $s_{i}=s_{i}^{\prime }$ or $%
j_{i}\in {\cal A}_{1}$, $s_{i}\preceq _{j_{i}}s_{i}^{\prime }$ and $%
s_{i}^{\prime }$ is acceptable for $s_{i}$, where we say that $s^{\prime }$
is {\em acceptable} for $s$ if there are
states $s_{h}$ and agents $i_{h}$, $h = 1, \ldots, k$, as described in
condition 2 for $s$, and $s^{\prime }=s_{i}$ for some $i\leq k$.
\end{enumerate}

We need to show that we can check whether $s$ seems consistent in time $O(|%
{\cal A}|2^{|\phi|})$.
The only difficulty is to determine, for given $s$ and
$s^{\prime}$, if $s^{\prime}$ is acceptable for $s$. It is clear that $%
k\leq|\phi|$, since we need at most one state and agent for each formula of
the form $\lnot E_{G}\psi\in s$. However, if we simply check each subgroup
of states containing $s^{\prime}$ and of agents containing $j$ where $%
s^{\prime }\in S_{j}^{1}$ that are of size $\leq|\phi|$ in the naive way,
this check will take time
at least $C(2^{|\phi|},|\phi|)C(|{\cal A}|,|\phi|)$
(where $C(n,k)$ is $n$ choose $k$),
which is
unacceptable for our desired time bounds. Instead, we proceed as follows.

Suppose that
$s' \in S_{i_1}^1$ and
$s\preceq_{i_{1}}s^{\prime}$. (If it is not the case that
$s\preceq_{i_{1}}s^{\prime}$,
then clearly $s^{\prime}$ is not acceptable for $s^{\prime}$.) Let $%
{F}(s,s^{\prime })$ consist of all formulas $E_{G}\psi$ such that
\begin{enumerate}
\item  $\lnot E_{G}\psi \in s$,

\item  $\lnot \exists t,i(i\in G\cap {\cal A}_{2}\,\wedge (s,t)\in {\cal K}%
_{i}\,\wedge \,\lnot \psi \in t)$, and

\item  $|A(s,s^{\prime },E_{G}\psi )|<|\phi |$, where $A(s,s^{\prime
},E_{G}\psi )=\{i\in G\cap {\cal A}_{1}:i=i_{1}\vee \,\exists t(s\preceq
_{i}t\,\wedge \,\lnot K_{i}\psi \in t)\}$.
\end{enumerate}
Intuitively, $F(s,s')$ consists of the potentially ``problematic''
formulas that may prevent $s'$ from being acceptable for $s$.

Let $T=\cup{_{E_G \psi \in F(s)}A(s,s}^{\prime},{E_{G}\psi)}$. Note that
$|T|<|\phi|^{2}$.
Suppose that $T=\{i_{1},\ldots,i_{N}\}$. We construct sets
$B_{1},\ldots,B_{N}$
of subsets of $F(s,s^{\prime})$ with the property that a set $X\in B_{k}$
iff there exist states $t_{1},\ldots,t_{k}$ such that $s\preceq_{i_{j}}t_{j}$
for $j=1,\ldots,k$ , $t_{1}=s^{\prime}$ and, for each formula $E_{G}\psi\in
X $, there exists a $j$ such that $\lnot K_{i_{j}}\psi\in t_{j}$.

Given a state $t\in S_{i}^{1}$, let $F_{t}(s,s^{\prime})=\{E_{G}\psi\in
{F}(s,s^{\prime}):\lnot K_{i}\psi\in t, \, i \in G\}$.  Intuitively,
$F_t(s,s')$ consists of the formulas in $F(s,s')$ that can be ``taken
care of'' by state $t$.
Let $B_{1}=\{F_{s^{%
\prime}}(s,s^{\prime})\}$. Suppose that we have defined
$B_{1},\ldots,B_{k}$. Let
$B_{k+1}=\{X\cup{F_{t}(s,s}^{\prime}{):X\in B_{k}\,\wedge s\preceq}_{i_{k+1}}%
{\,t\}}$. It is easy to check that $B_{k+1}$ has the required property.
Moreover, we can compute the sets $B_{1},\ldots,B_{N}$ in time $O(2^{cn})$.
To see this, note that since $|F(s,s^{\prime})|\leq|\phi|$, clearly $%
|B_{j}|\leq2^{|\phi|}$. Thus, given $B_{k}$, we can clearly compute $B_{k+1}$
in time $O(2^{cn})$ for some $c>0$. Since $N<|\phi|^{2}$, the result
follows. Finally, we claim that $s^{\prime}$ is acceptable for $s$ iff $%
{F}(s,s^{\prime})\in B_{N}$.

Clearly if $F(s,s^{\prime})\notin B_{N}$, then
it is almost immediate from the definition that
$s^{\prime}$ is not
acceptable for $s$. Conversely, if $F(s)\in B_{N}$, then there exist states $%
t_{1},\ldots,t_{N}$ such that $s^{\prime}=t_{1}$, $s\preceq_{i_{j}}t_{j}$
and, for each formula in $E_{G}\psi\in F(s)$, there exists $j$ such that $%
{\cal K}_{i_{j}}\psi\in t_{j}$. We clearly do not need all of these states
and agents; we just need at most one for each formula in $F(s,s^{\prime})$.
That is, there exists a set ${\cal A}^{\prime}$ of agents (contained in $%
\{i_{1},\ldots,i_{N}\}$) with $|{\cal A}^{\prime}|\leq|F(s,s^{\prime})|$ and
a state $u_{i}$ corresponding to each agent $i\in{\cal A}^{\prime}$
(contained in $\{t_{1},\ldots,t_{N}\}$) such that for each formula $%
E_{G}\psi\in F(s,s^{\prime})$, there exists an agent $i\in{\cal A}^{\prime}$
such that $s\preceq_{i}u_{i}$ and $\lnot K_{i}\psi\in u_{i}$. We now wish to
extend ${\cal A}^{\prime}$ to a set showing that $s^{\prime}$ is acceptable
for $s$. If we consider any $\lnot E_{G}\psi\in s$, either condition
2(a) is
satisfied or there is already an $i\in A^{\prime}$ satisfying 2(b) or $%
|A(s,s^{\prime},E_{G}\psi)|\geq|\phi|$. In the last case, it is immediate
that we can extend ${\cal A}^{\prime}$ to include 
an 
agent satisfying
2(b) for $E_{G}\psi$.

To show that this algorithm is correct, first suppose that $\phi$ is
satisfiable. In that case, $(M,s_{0})\models\phi$ for some structure $%
M=(S,\pi,\{{\cal K}_{i}^{\prime}:i\in{\cal A}\})\in{\cal M}^{rst}$. As for $%
{\cal M}^{rt}$, we can associate with each state $s\in S$ and $i\in{\cal A}%
_{1}$ the state $s_{i}^{\ast}$ in $S_{i}^{1}$ consisting of all the formulas
$\psi\in ESub_{i}(\phi)$ such that $(M,s)\models\psi$. It is easy to see
that if $(s,t)\in{\cal K}_{i}^{\prime}$ then $(s_{i}^{\ast },t_{i}^{\ast})\in%
{\cal K}_{i}$. Using this observation, a straightforward induction shows
that the states $s_{i}^{\ast}$ for $s\in S$ always seem consistent, and thus
are in $S^{j}$ for all $j$ and all $i\in{\cal A}_{1}$. Moreover, $%
\phi\in(s_{0})_{i}^{\ast}$ for all $i\in{\cal A}_{1}$. Thus, the algorithm
will declare that $\phi$ is satisfiable, as desired.

{F}or the converse, we need to show that if the algorithm declares that $\phi$
is satisfiable, then it is indeed satisfiable 
in ${\cal M}_{%
{\cal A}_{1}+{\cal A}_{2}}^{rst}$. We need to work a little harder than in
the previous proofs. Now we can no longer just view the object constructed
by our algorithm as the required structure. Rather, it serves as a
``blueprint'' for building the required structure.

Suppose that the algorithm terminates at stage $N$ with a state
$u\in S_{i_{u}}=S_{i_{u}}^{N}$ containing $\phi$.  Before we go on,
we make one observation that will prove useful in the sequel.
Notice that if $s \preceq_i s'$, then $E_G \psi \in s$ iff $E_G \psi \in
s'$ for $G \ne \{i\}$, and if
$j \in \A_2$, then $(s,t) \in \K_j$
iff $(s',t) \in \K_j$.
A {\em complete
state} is a vector $\vec{s} = 
(s^i: i \in \A_i \wedge s^i \in S^N_i)$
such
that
\begin{itemize}
\item $s^i \preceq_j s^j$ for all $i, j \in \A_1$ and
\item for every formula of the form $\neg E_G \psi \in \union_{i \in
\A_1} s^i$,  there exists an agent $j \in G$ and a state $t \in S^N$
such that $\neg \psi \in t$ and either
$j \in \A_1 \inter G$, $\neg K_j \psi \in s^j$, and $(s^j,t) \in
\K_j$ or $j \in \A_2$ and
$(s^i,t) \in \K_j$ for some $i \in \A_1$ (and hence $(s^i,t) \in \K_j$
for all $i \in \A_1$).
\end{itemize}
By consistency condition 2, every state $s \in S^N$ must be a
component of some (perhaps many) complete states.

Define
a structure $M^* = (S^*, \pi^*, \{\K_i^*: i \in \A_1 \union \A_2\}$ as
follows:
\begin{itemize}
\item $S^*$ consists of all complete states;
\item $\pi^*(\vec{s})(p) = {\bf true}$ iff $p \in \union_{i \in \A_1}
s^i$;
\item $(\vec{s},\vec{t}) \in \K_i^*$ for $i \in \A_1$ iff $s^i = t^i$
or $(s^i,t^i) \in \K_i$;
\item $(\vec{s},\vec{t}) \in \K_i^*$ for $i \in \A_2$ iff $(s^j,t^j)
\in \K_i$ for some $j \in \A_1$ (it is easy to check that if $(s^j,t^j)
\in \K_i$ for some $j \in \A_1$ then $(s^j,t^j) \in \K_j$ for all $j \in
\A_1$).
\end{itemize}

It is easy to check that $M^* \in \MPrsth$.  We now show that for
all $\psi \in \union_{i \in \A_i} \ESub^+_i(\phi)$, we have
$$
\mbox{$(M^*,\vec{s}) \sat \psi$ iff $\psi \in \union_{i \in \A_1} s^i$.}
$$
We proceed, as usual, by induction on the structure of $\psi$.
If $\psi$ is a primitive proposition, a conjunction, or a negation,
the argument is easy.  Suppose that $\psi$ is of the form $E_G\psi'$.
If $E_G \psi'
\in \union_{i \in \A_1} s^i$, then the construction of the $\K_j$
relations guarantees that $\psi' \in \union_{i \in 
\A_1}
t^i$ for all
$\vec{t} \in S^*$ such that $(\vec{s},\vec{t}) \in \K_j^*$ for some $j
\in G$.  Thus, by the induction hypothesis, we have
that $(M^*,\vec{s}) \sat E_G \psi'$.  For the converse, suppose that
$\neg
E_G \psi' \in \union_{i \in \A_1} s^i$.  Then from the definition of
complete
state and consistency condition 2, there must be some complete state
$\vec{t}$ and $j \in G$ such that $(\vec{s},\vec{t}) \in \K_j$ and $\neg
\psi' \in \union_{i \in \A_1} t^i$.

{F}inally, suppose that $\psi$ is of the form $C_G\psi'$.  If $C_G \psi'
\in \union_{i \in \A_1} s^i$ then, since $E_G (\psi' \land C_G \psi')$
must also be in $\union_{i \in \A_1} s^i$, an easy induction on the
length of the path shows that for every complete state $\vec{t}$
$G$-reachable from $\vec{s}$, we must have $\psi' \in \union_{i \in
\A_1} t^i$ so, by the induction hypothesis, we have
$(M^*,\vec{s}) \sat C_G \psi'$.  For the converse, suppose that $\neg
C_G \psi \in \union_{i \in \A_1} s^i$.  Then $\neg C_G \psi \in s^j$ for
some
(in fact, all) $j \in \A_1$.  If $G \inter \A_1 \ne \emptyset$, choose
$j \in G \inter \A_1$; otherwise, choose $j$ 
$\in G$ 
arbitrarily.
{F}rom consistency condition 3, it
easily follows that there exist complete states $\vec{s}_0, \ldots,
\vec{s}_k$ and $j_0, \ldots, j_{k-1} \in G$
such that $s_0^j = s^j$, $(\vec{s}_h,\vec{s}_{h+1}) \in \K_{j_h}^*$
$h = 0, \ldots, k-1$, and $\neg \psi' \in \union_{i \in \A_1} s_k^i$.
If $j \in \A_1$, then $(\vec{s}, \vec{s}_0) \in \K_j^*$; if $j \notin
\A_1$, then $j_0 \in \A_2$, and it follows from our initial observation
that $(\vec{s},\vec{s}_1) \in \K_{j_0}^*$.
In either case, $\vec{s}_k$ is $G$-reachable from $\vec{s}$, so
$(M^*,\vec{s}) 
\sat \neg C_G \psi'$, as desired. \eprf


We can now prove Theorem~\ref{complexity} for $\MPrst$.

\bigskip

\noindent
{\bf Proof of Theorem~\ref{complexity} for $\MPrst$:}
Again, the lower bound follows from standard results in \cite{HM2}.

{F}or the upper bound, suppose that we are given a formula $\phi$ such
that
$n
= |\phi|$.
We can compute the set $\E^m_{n}(\G_\phi)$
defined just before Lemma~\ref{Gmcount} in time $O(n^2 2^{cn})$,
using at most $n^2 2^n$ calls to the oracle $O_m$, just as we computed
$\E^1(\G_\phi)$.  Similarly, we can characterize the sets
$\H$ such that $(G,\H)$ is in
$\R(\G_\phi^m) = \R^m(\G_\phi)$
by a pair
$(\H',X)$, where $\H' \subseteq \G_\phi$ and $X$ is either $\emptyset$ or an
element of $\E^1_m(\G_\phi)$ and compute which of the pairs actually
represent sets in $\H$ such that $(G,\H) \in \R(\G_\phi^m)$ using
at most $2n(2^{2n} + 2^n)$ calls to the oracle $O_m$.
We cannot compute the individual elements of the sets $A_\H$
such that $|A_\H| \le m$, but it
does not matter.  It suffices that we know the cardinality of these
atoms (which our oracle will tell us).  We let $\A_1$ consist of the
agents $i_1^\H, \ldots, i_{|A_\H|}^\H$ for each $\H$ such that $|A_\H|
\le |\phi|$ ($i_1^\H, \ldots, i_{|A_\H|}^\H$ are just fresh names for
agents); let $\A_2$ consist of all $\H$ such that $|A_\H| > |\phi|$.

It is now straightforward to compute the formula $\phi^{\sigmarst}$ in
time $O(2^{cn})$ using $O(2^{cn})$ oracle calls.
We now
apply Proposition~\ref{transrst} and Theorem~\ref{decrst}, just as we
applied Proposition~\ref{trans} and Theorem~\ref{dec} in the case of
$\MP$.  \eprf

We now turn our attention to proving Theorem~\ref{complete} for
$\MPrst$.  Again, the basic structure is the same as for $\MP$ and
$\MPrt$.

\lem\label{replacerst} The mapping $\sigmarst$ (when viewed as a map
with domain $2^\A$) is injective on $\tG_\phi^n$.
\elem

Let $\fiveaxcp$ consist of the axioms in $\Taxcp$ (including E5)
together with E6 and every instance of K4 and K5 for $i \in \A_1$.
We write $\fiveaxcp
\vdash_\phi \psi$ if there is a proof of $\psi$ in $\fiveaxcp$ using
only the modal operators that appear in $\phi$ and $K_i$ for $i \in
\A_1$.

\lem\label{proofrst}  If $\A$ is finite and $\phi \in \LGC$ is
valid with respect to $\MPrsth$, then $\fiveaxcp
\vdash_\phi \phi$.   \elem

\prf The proof is similar in spirit to that of Lemma~\ref{proofrt} for
$\MPrt$, except that since we have a different definition of the $\K_i$
relations and of seeming
consistent, we must check that states eliminated under this definition
are inconsistent.  Again we must consider each of the three ways that a
state $s$ can be eliminated.

{F}irst, suppose that $s \in S^j$ and, for some $i \in \A_1$, there is no
$s'$ such that $s \preceq_i s'$.  As before, propositional reasoning
shows that
$\fiveaxcp \vdash_\phi \phi_s \dimp \lor_{\{s' \in S^1_i: s \preceq_i
s'\}} \phi_{s'}$.  Thus, it
easily follows that $\fiveaxcp \vdash_\phi \neg \phi_s$.

Next, suppose that $s\in S^{j}$ does not satisfy the second condition of
seeming consistent.
There must be a formula $\lnot E_{G}\psi \in s$ such that
\begin{enumerate}
\item  for all $i\in {\cal A}_{2}$ and all $t\in S^{j}$ such that $(s,t)\in
{\cal K}_{i}$, we have $\psi \in t$ and
\item  for all $i\in G\cap \A_{1}$, and all $t\in S^{j}$ such that $%
(s^{i},t)\in {\cal K}_{i}$, we have $\psi \in t$.
\end{enumerate}
Define an {\em extension of $s$\/} to be a vector $\vec{s%
}=(s^{i}:i\in \A_{1})$ of states, where $s\preceq _{i}s^{i}$. Let $EX(s)$ be
the set of all extensions of $s$. If $\vec{s}$ is an extension of $s$, let $%
\phi _{\vec{s}}$ be the conjunction over all $i\in \A_{1}$ of the formulas
in $\phi _{s^{i}}$. By straightforward propositional reasoning, we have $(%
{\rm S5}_\G^C)^{\A_1+\A_2}\vdash _{\phi }\phi _{s}\dimp\vee _{\vec{s}\in
EX(s)}\phi _{\vec{s}} $.
Thus, to show that
$\fiveaxcp \vdash_\phi \neg \phi_s$ if
$s$ is eliminated
by the second condition of seeming consistent, it suffices to show that
$({\rm S5}_\G^C)^{\A_1+\A%
_2}\vdash _{\phi }\lnot \phi _{\vec{s}}$ for each $\vec{s}\in EX(s)$.
This we do by showing that
$({\rm S5}_{G}^{C})^{\A_{1}+\A_{2}}\vdash_{\phi }\phi _{\vec{s}}\rimp
E_{G}\psi $
for each $\vec{s} \in EX(s)$.

So suppose that $\vec{s} \in EX(s)$.
The proof follows the lines of the analogous argument in the proof of Lemma~%
\ref{proofrt}. As before, it suffices to find, for each $i\in G$ and each
$t\in S_{i}^{j}$ with $\lnot \psi \in t$, a
set $G^{i,t}$ of agents containing $i$ such
that $({\rm S5}_{G}^{C})^{\A_{1}+\A_{2}}\vdash _{\phi }
\phi _{\vec{s}}\rimp %
E_{G^{i,t}}\lnot \phi _{t}$. For $i\in \A_{2}$, this follows as before if
the reason that $(s,t)\notin {\cal K}_{i}$ is that $s/\oKi \not\subseteq
t
$. If instead $t/\oKi \not\subseteq s$, then there is some $E_{G^{\prime
}}\theta \in t$ with $i\in G^{\prime }$ such that $\lnot \theta \in s$ and
so $\lnot \theta \in s^{i}$ for each $i$. Thus $({\rm S5}_{G}^{C})^{\A_{1}+\A%
_{2}}\vdash _{\phi }\phi _{\vec{s}}\rimp\lnot \theta $ and, by E6,
$({\rm S5}_{G}^{C})^{\A_{1}+\A_{2}}\vdash _{\phi }\lnot \theta \rimp
E_{G'}\lnot E_{G'}\theta $.  Since $E_{G'}\theta \in t$
we have that $({\rm S5}_{G}^{C})^{\A%
_{1}+\A_{2}}\vdash _{\phi }\phi _{\vec{s}}\rimp E_{G'}\lnot \phi _{t}$.
That is, we can take $G^{i,t} = G'$ in this case.

{F}or $i\in \A_{1}$, we show that $({\rm S5}_{G}^{C})^{\A_{1}+\A_{2}}\vdash
_{\phi }\phi _{\vec{s}}\rimp K_{i}\lnot \phi _{t}$
(so that we can take $G^{i,t} = \{i\}$).
By our assumption, $(s^{i},t)\notin \K_{i}$ for all $t \in S^i_j$.
Thus, if $t \in S^i_j$, there is some formula $\theta$ such that
either $K_i \theta \in s^i$
and $\neg K_i \theta \in t$ or $K_i \theta \in t$ and $\neg K_i \theta
\in s^i$.  Here we are
implicitly using the following facts: (1) if $E_{G'} \theta \in s$
for some
$G'$ such that $i \in G'$ then $K_i \theta \in s^i$, since $s^i \in
S_i^1$, and similarly for $t$, (2) if $K_i \theta \notin s$, then
$\neg K_i \theta \in s$, since $s^i \in S_i^j$, and similarly for $t$,
and (3) if $K_i \theta \in s^i$ then $\theta \in S$ since $(s,s) \in
\K_i$, and similarly for $t$.  If $K_i \theta \in s$ and $\neg K_i
\theta \in t$, it follows that $\fiveaxcp \vdash_\phi \phi_s \rimp
K_i \neg \phi_t$ just as in the case of $\fouraxcp$.  If $K_i \theta \in
t$ and $\neg K_i \theta \in s$, then by K5 we have
$\fiveaxcp \vdash_\phi \phi_s \rimp K_i \neg K_i \theta$
$\fiveaxcp \vdash_\phi \neg K_i \theta \rimp 
\neg\phi_t$.
The desired
result now follows by standard arguments.

We have now shown that for all $i \in G$ and $t \in S_j^i$ such that
$\psi
\in t$, there exists some set $G^{i,t}$ with $i \in G^{i,t}$ such that
$\fiveaxcp \vdash_\phi \phi_s \rimp E_{G^{i,t}} \neg \phi_t$.
We can now conclude that $\fiveaxcp \vdash_\phi \phi_s \rimp \neg E_G
\psi$ just as in the case of $\fouraxcp$, showing that $\phi_s$ is
inconsistent, as desired.

{F}inally, if $s \in S^j$ does not satisfy the third condition of
seeming consistent, the argument that $\fiveaxcp \vdash_\phi \neg
\phi_s$ is similar to that of Lemma~\ref{proof}. We replace $G$-reachability
by the existence of sequences as in
condition 3 in the definition of seeming consistent
in Theorem~\ref{decrst} and note that we have essentially already proved the
analogue of (\ref{eq2}) from Lemma~\ref{proof}. We leave the remaining
details to the reader. \eprf

\bigskip

\noindent
{\bf Proof of Theorem~\ref{complete} for $\MPrst$:} The proof follows as
for $\MPrt$ using the analogous lemmas proved above for $\MPrst$.
\eprf

\subsection{Dealing with $\M_\A^{\it elt}$}\label{sec:elt}

For $\MPelt$, we proceed much as for $\MPrst$.  There is one new
subtlety.  Consider the construction
in the proof of Proposition~\ref{transrst}, which uses $\sigmarst$.
Recall that
$\sigmarst(i)$ may be undefined for some $i$.  For such $i$, we defined
$\K_i$ to consist of all pairs $(s_\sigma,s_\sigma)$, making it
reflexive.  This approach will not work for $\MPelt$.  More precisely,
the analogue of Proposition~\ref{transrst} for $\MPelt$ will not hold
using this construction (even if we drop the reflexivity requirement).
For example, if $\phi = \neg p\land E_{G_1} p \land
E_{G_2} p$ and $G_1 \inter G_2 \ne \emptyset$, then $\phi^{\sigmarst}$
is satisfiable in $\MPelth$ but $\phi$ is not satisfied in the
structure $M'$ constructed in Proposition\ref{transrst}, since for all
$i \in G_1 \inter G_2$, the construction will make $\K_i$ reflexive.
We solve this problem by defining a mapping $\sigmaelt$ much like
$\sigmarst$, except that we ensure that $\sigmaelt$ is never undefined.

Let ${\cal B}$ be the set of
maximal subsets ${\cal T}$ of ${\cal G}_{\phi }$ such that $\cap {\cal T}%
\neq \emptyset $ and such that the corresponding atom over ${\cal
G}^{\phi }$, ${\cal A}_{{\cal T}}=(\cap {\cal T})\cap
(\cap_{G \in \G_\phi - \T\,} G)$
($=\cap {\cal T}$ by the maximality of ${\cal T}$), is not one of the
ones $%
{\cal A}_{{\cal H}}$ for ${\cal H}\in {\cal A}^{\phi ,rst}$. Let ${\cal
A}^{\phi ,elt}={\cal A}^{\phi ,rst}\cup {\cal B}$, ${\cal
A}_{1}={\cal B%
}\cup \{{\cal H}\in {\cal A}^{\phi ,rst}:|{\cal A}_{{\cal H}}|=1\}$, $%
{\cal A}_{2}={\cal A}^{\phi ,elt}-{\cal A}_{1}$. The definitions of
$\sigma
_{4}:{\cal A}\rightarrow {\cal A}^{\phi ,elt}$ and $\tau _{4}:{\cal A}^{\phi
,elt}\rightarrow 2^{{\cal A}}$ need some care. If $i\in {\cal A}_{{\cal H}}$
for some ${\cal H}\in {\cal A}^{\phi ,rst}$, let $\sigma _{4}(i)={\cal H}$
as before. Otherwise, choose ${\cal T}\in {\cal B}$ such that
${\cal T} \supseteq \{G\in {\cal G}_{\phi }: i\in G\}$ and let
$\sigma (i)={\cal T}$.
Note that, by construction, $\sigmaelt$ is defined for all $i$.
For ${\cal H}\in {\cal A}^{\phi ,rst}$, $\tau_{4}(\H)
=\cap \{\tG_{\phi}^{n}-\H\}$ as before. For ${\cal T}%
\in {\cal B}$, choose some $i_{{\cal T}}\in {\cal A}_{{\cal T}}$ (it
does not matter which) and set
$%
\tau _{4}({\cal T})=\{i_{{\cal T}}\}$.

\begin{proposition}\label{transelt}
$\phi $ is satisfiable in ${\cal M}_{{\cal A}}^{elt}$ iff $\phi ^{\sigma
_{4}}$ is satisfiable in ${\cal M}_{{\cal A}_{1}+{\cal A}_{2}}^{elt}$.
\end{proposition}

\prf
{F}irst suppose that $(M,s)\models \phi $, where $M\in {\cal M}^{elt}$.
We convert $M$
to $M' \in \MPelth$
as before by defining ${\cal K}_{{\cal I}}=\cup \{{\cal K}_{i}:i\in \tau
_{4}({\cal I})\}$ for ${\cal I}\in {\cal A}^{\phi ,elt}$.
To apply Proposition~\ref{trans}, we need to show that
$\union\{\tau(\I): \I \in \sigmaelt(G)\} = G$ for all $G \in \G_\phi$.
We know from the
analysis of the ${\cal M}^{rst}$ case that
$\cup \{\taurst(\H): \H \in \sigmarst(G)\} = G$ for all $G \in \G_\phi$.
Since $\sigma _{4}(G)\supseteq \sigma _{3}(G)$ and
$\tauelt(\H) = \taurst(\H)$ for $\H \in {\cal A}^{\phi,rst}$, we have
that $\cup \{\tauelt(\I):  \I \in \sigmaelt(G)\} = \cup \{\taurst(\H):
\H \in \sigmarst(G) \union \union\{\taurst(\I): \I \in \sigmaelt(G) -
\sigmarst(G)\}$.
It is clear from the definitions, however, that
if $\I \in \sigmaelt(G) - \sigmarst(G)$, then there exists some $i \in
G$ such that $\I = \sigmaelt(i)$ and $\sigmarst(i)$ is undefined.
Moreover, $\I = {\cal A}_{{\cal T}}\subseteq G$, so $\tauelt(\I) \in G$.
Thus, $\union\{\taurst(\I): \I \in \sigmaelt(G) - \sigmarst(G)\}
\subseteq G$, so
$\cup \{\tauelt(\I):  \I \in \sigmaelt(G)\} = \cup \{\taurst(\H):
\H \in \sigmarst(G)\} = G$, as desired.
Applying Proposition~\ref{trans}, we get that
to see that $(M^{\prime },s)\models \phi ^{\sigma _{4}}$.

It remains to verify that $M' \in \MPelth$.  For this, we need to
show that the ${\cal K}_{{\cal I}}$ relations for ${\cal I}\in \A_{1}$
are Euclidean, serial and transitive and that those in ${\cal A}_{2}$
are serial and secondarily reflexive. For the ones in ${\cal A}_{1}$,
note
that $\tau _{4}({\cal I})$ is a singleton and so the desired properties hold
since they hold for all agents in $M$. For the ones in ${\cal A}_{2}$, we
just note that the union of serial relations is serial and the union of
Euclidean relations is secondarily reflexive.

{F}or the other direction, we proceed much as in the proof of
Proposition~\ref{transrst}.
In addition to the concerns dealt with there for ${\cal M}^{rst}$,
our primary new one is to make sure that
the $\K_i$ relations for all agents
are serial. The problem
arises for those $i$ for which $\sigma _{3}(i)$ was undefined. The new
agents in ${\cal B}$ are used to deal with this problem.

We proceed much as in Proposition~\ref{transrst}, with two changes.
{F}irst, we replace the automatic forcing of reflexivity by forcing secondary
reflexivity for $\sigma _{3}(i)\in {\cal A}_{2}$. Second, we
modify the definition of the $\K_i$ relation in $M'$ as follows.
\begin{itemize}
\item If $\sigma_4(i) \in \A_1 \inter \A^{\phi,rst}$ then, as before,
$\K_i = \{(s_\sigma, t_{\sigma'}):
(s,t) \in\K_{\sigmarst(i)}, \sigma, \sigma' \in \Sigma\}$.
\item  If $\sigma _{4}(i)\in {\cal A}_{2}$ and $i\in G_{{\cal H},j}$,
then ${\cal K}_{i}=\{(s_{\sigma },t_{\sigma ^{\prime }}),(t_{\sigma'},
t_{\sigma ^{\prime }}):\sigma ^{\prime }=\sigma \cdot
((s,t),
i^j_\H)\}$.
\item  If $\sigma _{4}(i)={\cal T}\in {\cal B}$, then ${\cal K}%
_{i}=\{(s_{\sigma },t_{\sigma ^{\prime }}):(s,t)\in {\cal K}_{\sigma
_{4}(i)},\sigma ,\sigma ^{\prime }\in \Sigma \}$.
\end{itemize}

Now note that every relation ${\cal K}_{i}$ is Euclidean, serial and
transitive. For the ones corresponding to agents in ${\cal A}_{1}$ this is
immediate from the fact that the agents in ${\cal A}_{1}$ have these
properties. For those with $\sigma_{4}(i)\in{\cal A}_{2}$, seriality follows
from the fact that the agents in ${\cal A}_{2}$ are serial and the
construction. Transitivity and the Euclidean property follow from the
construction. In particular, if there is a ${\cal K}_{i}$ edge coming into
some $t_{\sigma}$ then there is none going out by construction except for
the one from $t_{\sigma}$ to itself.

The verification that $M^{\prime}$ satisfies $\phi$ now proceeds as in
Proposition~\ref{transrst}.
\eprf

\begin{theorem}
\label{decelt} If ${\cal A}={\cal A}_{1}+{\cal A}$ is finite and there is an
an algorithm for deciding if $i\in G$ for $G\in {\cal G}$ that runs in
time
linear in $|{\cal A}|$, then there is
a constant $c  > 0$ (independent of $|\A|$) and an algorithm
that, given a formula $%
\phi $ of ${\cal L}_{{\cal G}}^{C}$, decides if $\phi $ is satisfiable
in
${\cal M}_{{\cal A}_{1}+{\cal A}_{2}}^{elt}$ and runs
in time $O(|{\cal A}|2^{c|\phi |})$.
\end{theorem}

\prf
The argument here is like that for the ${\cal M}_{{\cal A}_{1}+{\cal A}%
_{2}}^{rst}$ case in Theorem~\ref{decrst}. We keep the definition of
${\cal K}_{i}$
for $i\in {\cal A}_{1}$ and, as we noted there, this makes these
relations
Euclidean and transitive. We change the definition of ${\cal K}_{i}$ for $%
i\in {\cal A}_{2}$ by putting $(s,t)$ in ${\cal K}_{i}$ iff
$s/\oKi \subseteq t$ and $t/\oKi \subseteq t$. This latter definition
clearly makes the ${\cal K}_{i}$ secondarily reflexive for $i\in {\cal A}_{2}
$. We ensure
seriality
by adding a clause to the definition of a state $s$
seeming consistent:

\begin{enumerate}
\item[4]  For every agent $i\in {\cal A}_{2}$ there is a state $t$ such that
$(s,t)\in {\cal K}_{i}$ and for every agent $i\in {\cal A}_{1}$ there are
states $s^{\prime }$ and $t$ such $s\preceq _{i}s^{\prime }$ and $(s^{\prime
},t)\in {\cal K}_{i}$.
\end{enumerate}

The proof now proceeds as before. \eprf

\bigskip

\noindent
{\bf Proof of Theorem~\ref{complexity} for $\MPelt$:}  The argument here
is essentially the same as for $\MPrst$. Just note that using the oracle
$O'$ we can determine the members of $\cal{B}$ within the appropriate
time bound and so compute $\phi^{\sigma_4}$ as required.
\eprf

We now turn our attention to proving Theorem~\ref{complete} for
$\MPelt$.  The basic structure is the same as for $\MPrst$.

\lem\label{replaceelt} The mapping $\sigmaelt$ (when viewed as a map
with domain $2^\A$) is injective on $\tG_\phi^n$.
\elem

Let $\Daxcp$ consist of the axioms in $\Kaxc$
together with K3, E4, E7, and every instance of K4 and K5 for $i \in
\A_1$. We write $\Daxcp
\vdash_\phi \psi$ if there is a proof of $\psi$ in $\Daxcp$ using
only the modal operators that appear in $\phi$ and $K_i$ for $i \in
\A_1$.

\lem\label{proofelt}  If $\A$ is finite and $\phi \in \LGC$ is
valid with respect to $\MPelth$, then $\Daxcp
\vdash_\phi \phi$.   \elem

\prf The proof is similar to that of Lemma~\ref{proofrst} for $\MPrst$.
Again we must check that all states eliminated in the construction are
provably inconsistent, but now using the axioms of $\Daxcp$ and the
modified
definition of the $\K_i$ relations, and dealing with the additional
clause in the definition of seeming consistent.

The argument for the first condition for seeming consistent
is the same as that for $\MPrst$.

Before dealing with the second condition, we prove a fact that will
also be useful in dealing with the fourth condition.  Let $T_i = \{t
\in S_i^j: (t,t) \in \K_i\}$.  It is easy to see that
\begin{equation}
\label{eqelt}
\mbox{if $t \notin T_i$, then $\Daxcp \vdash_\phi \phi_s \rimp E_G
\neg \phi_t$ for some $G$ such that $i \in G$.}
\end{equation}
For if $t \notin T_i$, then there exists $E_G \theta \in t$ such
that $i \in G$ and $\neg \theta \in t$.  But then $(E_G \theta \rimp
\theta) \rimp \neg \phi_t$ is propositionally valid (and so provable by
Prop).  Since $\Daxcp \vdash_\phi \phi_s \rimp E_G(E_G\theta \rimp
\theta)$, we can easily obtain (\ref{eqelt}) using (\ref{basic4}).

Now suppose that $s$ is
eliminated because it does not satisfy the second condition for seeming
consistent due to $E_{G}\psi$.
It again suffices to show that for each $i \in G$ and $t \in S_i^j$ such
that $\psi \in t$, there is a set $G^{i,t}$ of agents containing $i$ such
that $({\rm KD45}_{G}^{C})^{\A_{1}+\A_{2}}\vdash _{\phi }
\phi _{\vec{s}}\rimp  E_{G^{i,t}}\lnot \phi _{t}$.
First suppose $i \in \A_2$.
If $(s,t)\notin {\cal K}_{i}$ because $s/\oKi \not\subseteq t$ then
the argument given in Lemma~\ref{proof} works to get a $G^{i,t}$
as desired. If $s/\oKi$ $\subseteq t$ but $t/\oKi\not\subseteq t$ then
the existence of the required $G^{i,t}$ is immediate from (\ref{eqelt}).
Now suppose $i \in \A_1$
and $t$ is such that $(s^{i},t)\notin {\cal K}_{i}$. If
$s/K_{i}\not\subseteq t$, then there is some formula $\theta$ such that
$K_{i}\theta \in s^{i}$ and $\lnot \theta \in t$; it easily follows that
$\Daxcp \vdash_\phi \phi_{\vec{s}} \rimp K_i \neg \phi_t$, as required.
If $\{K_i \theta
:K_{i}\theta \in s\}\not\subseteq t$, then there is some $\theta $ such
that $%
K_{i}\theta \in s$ but $\lnot K_{i}\theta \in t$; the result now easily
follows using K4, just as in the argument for $\fouraxcp$.
If
both of these conditions hold (but still $(s^{i},t)\notin {\cal K}_{i}$),
then it must be that there is a $\theta $ with $K_{i}\theta \in t$ and $%
K_{i}\theta \notin s$. In this case $\lnot K_{i}\theta \in s$,
and the result follows using K5, just as in the argument for
$\fiveaxcp$.

The argument in the case that $s$ is
eliminated because it does not satisfy the third condition for seeming
consistent is the same as in the proof of Lemma~\ref{proofrst}.

Finally, suppose that $s$ does not satisfy
the new (fourth) condition of seeming consistent.  Then either
\begin{itemize}
\item  there is an $i\in {\cal A}_{2}$ for which there is no $t$ with $%
(s,t)\in {\cal K}_{i}$ or
\item  there is an $i\in {\cal A}_{1}$ for which there is no pair $s^{\prime
},t$ such that $s\preceq _{i}s^{\prime }$ and $(s^{\prime },t)\in {\cal K}%
_{i}$.
\end{itemize}

{F}or the first case,
for each $t \in T_i$, it must be the case that $s/\oKi \not\subseteq
t$, so that there must be some $G^{i,t}$ with $i \in G^{i,t}$ such that
$\Daxcp \vdash \phi_s \rimp E_{G^{i,t}} \neg \phi_t$, as usual.  By
(\ref{eqelt}), for each $t \notin T_i$, there is some $G^{i,t}$ with $i
\in G^{i,t}$ such that
$\Daxcp \vdash_\phi \phi_s \rimp E_{G^{i,t}} \neg \phi_t$.  Thus,
$\Daxcp \vdash_\phi \phi_s \rimp \land_{t \in S^j} E_{G^{i,t}} \neg
\phi_t$. But since
$\Daxcp \vdash_\phi \neg (\land_{t \in S^j} \neg \phi_t)$ by 
induction and
propositional reasoning, it follows from E7 that
$\Daxcp \vdash_\phi \neg (\land_{t \in S^j} E_{G^{i,t}} \neg
\phi_t)$.  Thus we get
$\Daxcp \vdash_\phi \neg \phi_s$, as desired. 

{F}or the second case, we know as in the proof of Lemma \ref{proofrst}
that $\phi _{s}$ is provably equivalent to the disjunction of $\phi
_{s^{\prime }}$ for those $s^{\prime }$ such that $s\preceq _{i}s^{\prime }$
and similarly for any $t$. Thus to prove
$\Daxcp \vdash_\phi \lnot \phi _{s}$ it suffices
to prove
$\Daxcp \vdash_\phi \lnot \phi _{s^{\prime }}$ for every $s^{\prime }\in
S_{i}^{j}$ such
that $s\preceq _{i}s^{\prime }$. For each such $s^{\prime }$ we know that
there is no $t^{\prime }\in S_{i}^{j}$ such that $(s^{\prime },t^{\prime
})\in {\cal K}_{i}$.
Given $s'$, if $t' \in S_i^j$ and 
$(s',t') 
\notin \K_i$, then the same
argument as in the proof of Lemma~\ref{proofrst} shows that
$\Daxcp \vdash_\phi \phi_{s'} \rimp K_i \neg \phi_{t'}$, since the
$\K_i$
relations are defined the same way for agents in $\A_1$ in both 
the 
$\MPelt$
and $\MPrst$ 
cases,
and the proof in Lemma~\ref{proofrst} used only
axioms K4 and K5 (as well as Prop, K1, and MP), and these axioms are in
both $\fiveaxcp$ and $\Daxcp$.

By (\ref{basic3}), we have that $\Daxcp \vdash \phi_{s'} \rimp
K_i(\land_{t' \in S_i^j} \neg \phi_{t'})$.  Since $\Daxcp \vdash_\phi
(\land_{t' \in S_i^j} \neg \phi_{t'}) \rimp \false$ by 
induction and 
propositional
reasoning, we conclude that $\Daxcp 
\linebreak[2]
\vdash_\phi \phi_{s'} \rimp K_i
\false$.  Now using K3, we get $\Daxcp \vdash_\phi \neg \phi_{s'}$, as
desired. \eprf

\bigskip

\noindent
{\bf Proof of Theorem~\ref{complete} for $\MPelt$:} The proof follows as
for $\MPrt$ using the analogous lemmas proved above for $\MPelt$.
We must just show that E7 is derivable from the other axioms in $\Daxc$.
Suppose that $i \in G_1 \inter \ldots G_k$.  Then, using E1, $\Daxc
\vdash E_{G_1}
\phi_1 \land \ldots \land E_{G_k} \phi_k \rimp K_i \phi_1 \land \ldots
\land K_i \phi_k$.  By (\ref{basic3}), we have
$\Daxc \vdash K_i \phi_1 \land \ldots \land K_i \phi_k \rimp K_i(\phi_1
\land \ldots \land \phi_k)$.  Thus, $\Daxc \vdash \neg K_i(\phi_1 \land
\ldots \land \phi_k) \rimp 
\neg (E_{G_1} \phi_1 \land \ldots \land E_{G_k} \phi_k)$.
It thus suffices to show that in $\Daxc$, from $\neg(\phi_1
\land \ldots \land \phi_k)$ we can infer $\neg K_i(\phi_ \land \ldots
\land \phi_k)$.  But since $\neg (\phi_1 \land \ldots \land \phi_k)$ is
equivalent to $(\phi_1 \land \ldots \phi_k) \rimp \false$, this follows
easily using (\ref{basic4}) and K3.
\eprf

\subsection{The complexity of querying the oracles}\label{oracle}
Up to now we have assumed that we are charged one for each query to an
oracle.  In this section, we reconsider our results, trying to take into
account more explicitly the cost of the oracle queries.

Let $f(m,k)$ be the worst-case time complexity
of deciding whether a set with description $G \in \hG_\A^m$
such that $l(G) \le k$ has cardinality greater $m'\le m$ (where we take
the worst case over all $G \in \hG_\A^m$ such that $l(G) \le
k$ and over all $m' \le m$).
Let $g(k)$ to be the worst-case complexity of deciding if
$G_1 \inter \ldots \inter G_k = \emptyset$ for $G_1, \ldots, G_k \in
\G_\A$.
We take $f(m,k)$ (\respc $g(k)$) to be $\infty$ if these questions are
undecidable.
We can think of $f(m,k)$ (\respc $g(k)$) as the worst-case cost of
querying the oracle $O_m$ (\respc $O'$)
on a set
with a
description of length $\le k$.

Using these definitions,
we can sharpen Theorem~\ref{complexity} as follows.
\thm\label{complexity1} There is a constant $c > 0$ and an algorithm
that decides if
a formula $\phi \in \LGC$ is satisfiable in
$\MP$ (\respc
$\MPr$, $\MPrt$, $\MPrst$, $\MPelt$) and runs in time
$2^{c|\phi|}f(0,|\phi|)$
(\respc
$2^{c|\phi|}f(0,|\phi|)$,
$2^{c|\phi|}f(1,2^{c|\phi|^2})$,
$2^{c|\phi|}f(|\phi|,2^{c|\phi|^2})$,
$2^{c|\phi|}(f(|\phi|,2^{c|\phi|^2}) + g(|\phi|))$)
Moreover, if $\G$
contains
a subset with at least two elements, then there exists a constant $d >
0$ such that every algorithm for deciding the satisfiability of formulas
in $\MP$ (\respc $\MPr$, $\MPrt$, $\MPrst$, $\MPelt$) runs in time at
least
$\max(2^{d|\phi|},f(0,d|\phi|))$ (\respc
($\max(2^{d|\phi|},f(0,d|\phi|))$,
$\max(2^{d|\phi|},f(1,d|\phi|))$,
$\max(2^{d|\phi|},f(d|\phi|,d|\phi|))$,
$\max(2^{d|\phi|},f(d|\phi|,d|\phi|),g(d|\phi|))$)
for infinitely many formulas $\phi$.
\ethm

\prf The upper bound is almost immediate from the proof of
Theorem~\ref{complexity}.  The only point that needs discussion is the
second argument---$2^{c|\phi|^2}$---of $f$ in the cases $\MPrt$,
$\MPrst$, and $\MPelt$.  This follows from Lemma~\ref{Gmcount}.
An easy induction on $i$ shows that the sets in
the set
$\E_i^{|\phi|}$
constructed just before Lemma~\ref{Gmcount} have
description length at most
$\le 2^{2i|\phi|}$
(using the
fact that
$|\E_i^{|\phi|}|
\le 2^{|\phi|}$ for all $i$).  Thus, all the sets that
we need to deal with have description length $\le 2^{2|\phi|^2}$, since
they are all in
$\E_{|\phi|}^{|\phi|}$,
by
Lemma~\ref{Gmcount}(e).

The lower bound is immediate from the results of \cite{HM2} and
Proposition~\ref{reduction}.  \eprf

Note that if $f_0(k) = f(0,k)$ is
{\em well behaved}, in that
there exist $c'$, $k_0$ such that $f_0(k) \le
2^{c'k}$ for all $k \ge k_0$ or $f_0(k) \ge 2^{c'k}$ for all $k \ge
k_0$, then it is easy to see that there is some
$c'' > 0$ such that
$2^{c|\phi|}f(0,|\phi|) \le
\max(2^{c''|\phi|},c''f(0,|\phi|))$.
Thus, if $f_0$ is well behaved, then the lower and upper bounds
of Theorem~\ref{complexity} match, and we have tight bounds in the case
of $\MP$ and $\MPr$.  This is not the case for $\MPrt$, $\MPrst$, and
$\MPelt$, because the sets that arise have exponential-length
descriptions.

Do we really have to answer queries of about such complicated formulas
if we are to deal with $\MPrt$, $\MPrst$, and $\MPelt$?  To some extent,
this is an artifact of our insistence that the sets be described using
union and set difference.  In fact, all the sets that we need to
consult the oracle about in our
algorithm are atoms, and so have very simple descriptions
($O(|\phi|)$) if we are allowed to used intersections and complementation.
Indeed, suppose that we define an ordering $\prec$ on
atoms such that $A_\H \prec A_{\H'}$ if $\H \supset \H'$.  It follows
easily
from Lemma~\ref{unique} and Lemma~\ref{K} that in order to compute
$\sigmart(G)$ (\respc $\sigmarst(G)$, $\sigmaelt(G)$), we start by
considering all atoms $A_\H$ such that $G$ appears positively in $A_\H$
and all other sets in $\G_\phi$ appear negatively; we then need to check
whether $|A_\H| > 0$ and $|A_\H| > 1$ (\respc $|A_\H| > 0$, \ldots, and
$|A_\H| > |\phi|$) only for those atoms $A_\H$ such that for all $\H'
\prec \H$, we have $|A_{\H'}| \le 1$ (\respc $|A_{\H'}| \le |\phi|$).
(In addition, in the case of $\sigmaelt$, we have also have to check
whether $G_1 \inter \ldots
\inter
G_k = \emptyset$, but again, these are sets
with simple descriptions if we allow intersection.)
Thus, as long as we can check the required properties of sets
described in terms of intersection and complementation relatively
efficiently, then the queries to the oracle pose no problem.
Unfortunately, the bounds in Proposition~\ref{reduction} depends on the
descriptions involving only set difference and union, so we cannot get
tight bounds for Theorem~\ref{complexity} (at least, with our current
techniques) using descriptions that involve intersection and
complementation.
It remains an open question whether we can get tight bounds in all cases
taking into account the cost of querying the oracle.

\section{Conclusions}\label{discussion}
We have characterized the complexity of satisfiability for epistemic
logics when the set of agents is infinite.  Our results emphasize the
importance of how the sets of agents are described and
provide new information
even in the case where the sets involved are finite.

In this paper we have focused on a language that has operators $E_G$ and
$C_G$.  There are two interesting directions to consider extending our
results.
\begin{itemize}
\item We could restrict the language so that it has only $E_G$
operators.  If the set of agents is finite (and all sets $G$ are
presented in such a way that it is easy to check if $i \in G$), then
there are well-known results that show the complexity of the decision
problem in this case is PSPACE complete \cite{HM2}.  However, again,
this result counts $E_G$ as having length $|G|$.  Although we have not
checked details, it seems relatively straightforward to combine the
techniques of \cite{HM2} with those presented here to get PSPACE
completeness for $\LGE$, taking $E_G$ to have length 1, using the same
types of oracle calls as in Theorem~\ref{complexity}.
(Note that
Proposition~\ref{reduction} applies to the language $\LGE$; we did not
use the $C_G$ operators in this proof.)
\item We could add the distributed knowledge operator $D_G$ to the
language \cite{FHMV,FHV2,HM2}.  Roughly speaking, $\phi$ is distributed
knowledge if the agents could figure out that $\phi$ is true by pooling
their knowledge together.  Formally, we have
\begin{quote}
$(M,s) \sat D_G \phi$ if
$(M,t) \sat \phi$ for all $t \in \inter_{i \in G} \K_i(s)$.
\end{quote}
It is known that if $\A$ is finite (and there is no difficulty in
telling if $i \in G$), then adding $D_G$ to the language poses no
essential new difficulties \cite{FHMV,HM2}.  We can get a complete
axiomatization, the satisfiability problem for the language with $D_G$
and $E_G$ operators is PSPACE complete, and once we add common
knowledge, the satisfiability problem becomes exponential-time complete.
Once we allow infinitely many agents, adding $D_G$ introduces new
subtleties.
For example, 
even if we place no assumptions on the $\K_i$ relations,
once we have both $E_G$ and $D_G$ in the language, we need to be able
to distinguish between sets of 
cardinality one and those with larger cardinality since 
$E_Gp \dimp D_Gp$ is valid if and only if $G$ is a singleton. New 
issues also arise 
once we make further assumptions about the $\K_i$ relations
because different 
properties are preserved for the new agents, say 
$K_{\A^D}$ and $K_{\A^E}$, which are to be added on as in 
Proposition~\ref{Shore} to represent $D_\A$ and $E_\A$, respectively. 
Intuitively, 
$\K_{\A^E}$ corresponds to the union of the 
relations $K_i$ for $i \in G$
while $\K_{\A^D}$ corresponds to their intersection. Thus, while both  
$K_{\A^D}$ and $K_{\A^E}$ inherit reflexivity and symmetry from the
$K_i$ relations,
$K_{\A^D}$ inherits transitivity and the Euclidean property 
while $K_{\A^E}$ does not.  
There are also additional 
relations between these agents that must be taken into account. 
Examples in S4 and S5 include $K_{\A^E}\phi \rimp K_{\A^D} \phi$, 
$K_{\A^E}K_{\A^D}\phi \rimp K_{\A^E}\phi$ and 
$K_{\A^D}K_{\A^E}\phi \rimp K_{\A^E}\phi$. 

These are issues for future work.
\end{itemize}

\bibliographystyle{alpha}
\bibliography{z,refs,joe}

\end{document}